\documentclass[fleqn]{elsart}
\usepackage{graphicx,color}
\usepackage{amsmath,amssymb}
\usepackage{slashbox}
\usepackage{young}
\newcommand{\Slash}[1]{{\ooalign{\hfil/\hfil\crcr$#1$}}}
\newcommand{\tr}{{\rm tr}}
\newcommand{\Tr}{{\rm Tr}}
\newcommand{\Ln}{{\rm Ln}}

\newcommand{\Nc}{N_{\rm c}}
\newcommand{\Nf}{N_{\rm f}}

\newcommand{\lqcd}{\Lambda_{\rm QCD}}

\newcommand{\vp}{\vec{p}}

\newcommand{\vr}{\vec{r}}

\newcommand{\vx}{\vec{x}}
\newcommand{\vy}{\vec{y}}

\newcommand{\vw}{\vec{w}}
\newcommand{\vu}{\vec{u}}

\newcommand{\la}{\langle}
\newcommand{\ra}{\rangle}

\newcommand{\calL}{\mathcal{L}}

\newcommand{\calS}{\mathcal{S}}

\newcommand{\calD}{\mathcal{D}}
\newcommand{\calN}{\mathcal{N}}

\newcommand{\nd}{n_{\rm d}}
\newcommand{\rmd}{\mathrm{d}}
\newcommand{\rmi}{\mathrm{i}}
\newcommand{\rme}{\mathrm{e}}

\newcommand{\up}{\uparrow}
\newcommand{\down}{\downarrow}
\newcommand{\udown}{u\!\downarrow}
\newcommand{\uup}{u\!\uparrow}
\newcommand{\dup}{d\!\uparrow}
\newcommand{\ddown}{d\!\downarrow}
\newcommand{\rMS}{ {\rm MS} }
\newcommand{\gA}{g_A^{ {\rm val} } }
\newcommand{\cal}{\mathcal}

\begin{document}
\begin{frontmatter}
\title{Can the nucleon axial charge be $O(\Nc^0)$?}
\author{Toru Kojo}
\address{Faculty of Physics, University of Bielefeld,
Bielefeld, Germany}
\begin{abstract}
%
%
%
%
The nucleon self-energy and its relation to the 
nucleon axial charge $g_A$
are discussed at large $\Nc$.
The energy is compared for the hedgehog,
conventional, and recently proposed dichotomous 
nucleon wavefunctions which give
different values for $g_A$.
We consider their energies at both perturbative
and non-perturbative levels.
In perturbative estimates,
we take into account the pion exchanges among quarks
up to the third orders of axial charge vertices,
including the many-body forces such as the Wess-Zumino terms.
It turns out that
the perturbative pion exchanges among valence quarks
give the same leading $\Nc$ contributions for three wavefunctions,
while their mass differences are $O(\lqcd)$.
The signs of splittings flip for different orders
of the axial charge vertices,
so it is hard to conclude which one is the most energetically
favored.
For non-perturbative estimates involving
the modification of quark bases,
we use the chiral quark soliton model as an illustration.
With the hedgehog quark wavefunctions with $g_A$ of $O(\Nc)$,
we investigate whether solutions with coherent pions
are energetically favored.
Again it is hard to give decisive conclusions,
but it is possible that adding the confining effects
disfavors the solution with the coherent pions,
making a pion cloud around a nucleon
quantum rather than coherent.
The nuclear matter at large $\Nc$ is also discussed
in light of the value of $g_A$.
\end{abstract}
\end{frontmatter}
%


\section{Introduction}

Recently we discussed the phenomenological problems
for nucleons at large $\Nc$,
which are caused by the coherent pions
surrounding nucleons \cite{Hidaka:2010ph}.
The source of a coherent pion cloud is
the large nucleon axial charge $g_A$ of $O(\Nc)$,
whose value depends upon
quark wavefunctions inside a nucleon.

The problems related to large $g_A$ or coherent pions
manifestly appear in the pion exchange part 
of the nuclear potential.
Consider dilute nuclear matter
where the pion exchange is the most important.
For the nucleon-nucleon potential,
the overall strength of the pion exchange is proportional to 
the square of the vertex, $(g_A/f_\pi )^2 \sim \Nc$.
The kinetic energy of nucleons are 
at most $\sim \vp^2/M_N \sim \lqcd/\Nc$,
so the kinetic energy is much smaller than 
the potential energy.
It means that at large $\Nc$,
nucleons localize at the potential minima,
forming a crystal even at dilute regime
\cite{Kutschera:1984zm,Klebanov:1985qi,Kugler:1989uc,Nawa:2008uv}.
In reality, however, dilute nuclear matter is 
like a liquid where the potential and kinematic effects
are equally important,
and its binding energy is $O(1-10)$ MeV,
hardly recognizable from
the QCD scale $\lqcd \sim 200$ MeV.
For this qualitative discrepancy from the $\Nc=3$ world,
usually the large $\Nc$ results are not
regarded as useful guidelines for studies of a nuclear matter.

This situation, however, might be altered
if the value of $g_A$ were not $O(\Nc)$ but $O(1)$.
With $g_A \sim 1$, 
both the potential and kinetic energy become comparable,
$\sim \lqcd/\Nc\sim 60$ MeV at $\Nc=3$,
so it is much easier to expect the $O(1-10)$ MeV binding energy,
by cancelling out the energetic cost and gain
at the same order.
Then the large binding energy problem
could be solved, or at least be largely reduced.
In this picture of small $g_A$,
the long distance part of nucleon potentials
should be described not by coherent pions
but by a few quantum pions,
and in fact, the latter
has been a description in conventional nuclear 
physics \cite{Yukawa35,Lacombe:1980dr,Machleidt:1989tm}.
Motivated by this observation,
in Ref.\cite{Hidaka:2010ph}
we started our attempt to construct
the ground state nucleon wavefunction
with $g_A \sim 1$
in case of the $SU(\Nc)$ gauge theory. 
In this paper,
we shall continue our discussions about
a nucleon wavefunction with small $g_A$.

This reduction of $g_A$ itself 
does not solve all of the problems, however.
We must handle
a large attractive force of $O(\Nc)$
at intermediate distance as well.
It comes from the exchange of the $\sigma$-meson
whose mass is supposed to be
sensitive to $\Nc$ 
in contrast to other mesons 
\cite{Pelaez:2003dy,Kojo:2008ih}.
We imagine that at large $\Nc$, the
range of $\sigma$ is comparable to the range of 
the strong repulsive force originating from the $\omega$ meson,
and the spatial size of the attractive pocket becomes very small.
Then the binding energies of nucleons
would arise not from the leading $\Nc$ contributions,
but from the $1/\Nc$ corrections. 
This aspect will be discussed elsewhere,
but in what follows,  the longest range
part of the interaction must be first resolved
before arguing the forces at intermediate distance.
Therefore, as a first step,
we focus on the pion problem.

The source of the axial charge $g_A$ is
quark dynamics inside a nucleon. 
In this work, we will consider the valence quark
contributions to the axial charge, $g_A^{ {\rm val} }$,
that is estimated from the valence nucleon wavefunction.
As a theoretical guideline, in the following 
we consider problems in the context
of the constituent quark picture, including the meson exchanges.
If we include only the pions,
the description we shall use is 
analogous to the frameworks
of Manohar and Georgi \cite{Manohar:1983md},
Weinberg \cite{Weinberg:2010bq},
and Glozman and Riska \cite{Glozman:1995fu}.
Following a study of Weinberg \cite{Weinberg:1990xm}, 
the quark axial charge is assumed to be $1$.

For considerations of the axial charge,
it is convenient to use the $SU(4)$ spin-flavor
generators \cite{Gervais:1983wq,Dashen:1993jt}, formed by
isospin, spin and axial charge operators:
\begin{equation}
\tau_a= \sum_q \tau_a^{(q)}\,,~
\sigma_i = \sum_q \sigma_i^{(q)}\,,~
R_{ai} = \sum_q R_{ia}^{(q)} = \sum_q \tau_a^{(q)} \sigma_i^{(q)} \,, 
\end{equation}
%
where $q$ means the operator acts on the $q$-th quark.
Here we stress that
we will {\it not} require the Hamiltonian and
its eigenstates to be $SU(4)$ symmetric,
and in fact, nucleons are not eigenstates of 
axial charge generators.
The use of the $SU(4)$ generators
is {\it not} mandatory step for our discussions.
Nevertheless the use of the $SU(4)$ makes it easier
to get several qualitative insights
about quantities related to the axial charge operators. 
For instance,
the diagonal axial charge of the proton with spin up is given by
\begin{equation}
\gA = \la p \! \up \!| R_{33} | p \!\up \ra \,,
\end{equation}
for the valence quark part. 
When we write $|p\!\up\ra$, it means
the valence quark part for which we do not 
include extra contributions from induced pion clouds, etc., 
unless otherwise stated.

Below we will consider three types of the
valence baryon wavefunctions in order.
The hedgehog, conventional,
and recently proposed dichotomous 
wavefunctions \cite{Hidaka:2010ph}.
They provide different values of $\gA$:
$-\Nc$, $\pm (\Nc+2)/3$, and $\pm 1$, respectively.

In many studies, large $\Nc$ baryons are discussed
starting with the baryon wavefunction of 
the hedgehog type \cite{Manohar:1984ys}.
The hedgehog state is a mixture of different spins and isospins
but has a definite and very large axial charge, 
$g_A^{ {\rm val} }=-\Nc$,
that generates a large pion field.
It allows us to apply the mean field or coherent picture of pions
which is an underlying basis for the 
Skyrme type models 
\cite{Skyrme:1961vq,Adkins:1983ya,Zahed:1986qz},
its holographic version 
\cite{Sakai:2004cn,Hong:2007kx,Nawa:2006gv},
or chiral quark soliton models 
\cite{Kahana:1984dx,Diakonov:1987ty,Reinhardt:1988fz,Alkofer:1994ph,Christov:1995vm}.
The states with definite isospin and spin
can be obtained by collectively rotating
the hedgehog wavefunction and the pion cloud.
Therefore nucleons constructed in this way 
(we call them hedgehog nucleons)
accompany a coherent pion cloud.

A more theoretically sound way to compute 
nucleon properties is to always keep the quantum numbers
during calculations,
although such computations are technically much more involved.
It is nontrivial whether such a nucleon
involves a coherent pion cloud or not.
This is because 
once we fix the isospin and spin of a nucleon state,
it is inevitably a mixture of different
axial charge states (See discussions later).

In the conventional nucleon wavefunction,
all quarks occupy the lowest $S$-wave orbit,
and the spin-flavor wavefunction is maximally symmetrized
to satisfy the Fermi statistics.
The conventional nucleon wavefunction
produces $g_A^{ {\rm val} } = \pm (\Nc+2)/3$
at odd $\Nc$, while at even $\Nc$
it becomes zero \cite{Hidaka:2010ph} 
(here the ``nucleon'' at even $\Nc$ simply means 
the lowest isospin and spin state).
This big sensitivity of $g_A^{ {\rm val} }$ to whether $\Nc$ is odd or even
is a consequence that a nucleon wavefunction
is a superposition of different axial charge states.
In particular, for an even $\Nc$ ``nucleon'',
the contributions from different axial charge states
completely cancel out one another.

In the dichotomous wavefunction at odd $\Nc$,
we place $\Nc-1$ quarks in the lowest $S$-wave orbit
in such a way that they have zero 
isospin and spin.
Then the remaining quark carries the same quantum number 
as nucleons, and we place it in a spatial orbit different from
the other $\Nc-1$ quarks.
As seen from the computation for the $\Nc$ even ``nucleon'',
the contributions to $\gA$ from the $\Nc-1$ quarks
cancel out one another, and $\gA$ is saturated solely by
a single quark.
It means that $\gA$ is $1$.
An obvious question, however, is how
to assure that the dichotomous state
is energetically favored compared to the conventional one.
To place one quark in an orbit different from
the lowest $S$-wave orbit,
we have to pay an energy penalty of $O(\lqcd)$.

Obviously which states appear as the ground state
depends on the quark dynamics inside of the baryons.
The purpose of this paper 
is to give the arguments on the energetic differences
originating from the axial charge operators.

Our philosophy has several similarities 
with the chiral bag model as a hybrid model
of quarks and pions \cite{Brown:1979ui,Theberge:1980ye,Hosaka:1996ee}.
Inside of a confining bag,
there are quarks creating the axial charge
as a source of pions.
The strength of the source
determines the size of the self-energy from the pion loops
which modify the mass of the valence nucleon considerably.
There are several versions depending on how 
we choose the size of the bag.
But according to the Cheshire cat 
principle \cite{Nadkarni:1985dn},
it is just a matter of practice to make
the computation of quarks and pions (with topological
quark number)
more tractable.
Phrasing our treatment in this context, 
for the quark part 
we take a constituent quark model 
with the bag radius large enough to develop
the constituent quark mass, and apply
the non-relativistic framework to
the computation of the valence axial charge.

We will argue the energy dependence of a nucleon
on $g_A$ in perturbative and non-perturbative contexts.
Unfortunately,
it turns out that  in both regimes, 
it is very hard to derive a definite conclusion
about which one is the lightest state.

In a perturbative context,
we consider pion exchanges among quarks inside a nucleon,
which are supposed to be strongly correlated to the
axial charge of a nucleon.
The summation of their contributions
depend on the spin-flavor wavefunctions,
and are computed group theoretically.
We evaluated the self-energy of baryons
up to the terms quadratic and cubic 
in the axial charge operators.
The Wess-Zumino vertices 
\cite{Wess:1971yu,Kaymakcalan:1983qq,Fujiwara:1984mp}
are included.
It turns out that the many-body forces
or higher orders of axial charge operators
appear at the same order of the leading $\Nc$,
so that the convergence is not guaranteed.

We will argue that
at the level of perturbative evaluations,
the leading $\Nc$ contributions for
hedgehog, conventional, and dichotomous wavefunctions
are energetically degenerate.
The pion exchange contributions for these wavefunctions
differ by at most $O(\lqcd)$, at each order of the axial charge operators.
So three wavefunctions differ in energy by $O(\lqcd)$
unless there are subtle cancellations among
different orders of axial charge operators.

The computations based upon the perturbative pions
are certainly insufficient, especially
when we argue the nucleon with a large axial charge.
When the strength of the pion field becomes very large,
such a large pion field may drill a hole 
in a media of the chiral condensate, costing energy.
Accordingly the constituent quark wavefunctions are modified as well.
In this regime,
we must take the non-perturbative effects into account
at the stage determining the quark bases.

This kind of contribution is considered
in the chiral quark soliton model including a full evaluation of 
the quark determinant \cite{Diakonov:1987ty,Reinhardt:1988fz,Alkofer:1994ph,Christov:1995vm},
provided that the background coherent pion field
is of the hedgehog type.
In this picture, 
the production of the coherent pion cloud costs energy of
$O(\Nc)$, 
while  quarks acquire energetic gains of $O(\Nc)$,
by being bound to the pion cloud.
The optimized configuration of pions and quarks 
is determined by
the energetic balance between these $O(\Nc)$ contributions.

We will see that for the hedgehog wavefunctions of quarks,
it is justified to take the pion fields to be
static and classical.
Within a context of the chiral quark soliton picture,
we give arguments why we expect that
the hedgehog wavefunction is not energetically favored
if the confining effects are included.
The point is that the confining forces
restrict the spatial size of quark wavefunctions,
preventing deeply bound quarks to a pion cloud. 
On the other hand, 
for the conventional or dichotomous nucleon wavefunctions,
we do not know how to take the appropriate mean field for pions, 
so we could not show the non-perturbative expressions of
energies for these two wavefunctions.

In this paper, we discuss only two flavor cases,
close to the chiral limit.
Our metric is $g_{\mu \nu}= {\rm diag}(1,-1,-1,-1)$.

This paper is organized as follows.
Sec.\ref{mesonloop} is devoted to 
some preparation of basics which are used in later sections.
We start with counting of the quark-meson coupling 
and classify its total contribution 
based on the spin-flavor wavefunctions of nucleons.
Then we briefly review basic relations of the $SU(4)$ generators
to the extent necessary for our arguments.
In Sec.\ref{Wavefunctions},
we show the construction of
the hedgehog, conventional, and dichotomous wavefunctions,
and compute their axial charges.
In Sec.\ref{Pert},
we compare the energies of three wavefunctions
at the perturbative level.
In Sec.\ref{Nonpert},
we give non-perturbative considerations within 
a context of the chiral quark soliton picture.
Sec.\ref{Discussion}
is devoted to summary.

\section{Meson exchanges for
the baryon self-energy:
Preparations}\label{mesonloop}

In this section,
we first argue how
the meson exchanges among quarks
contribute to the baryon mass.
Especially we will focus on the meson exchange
in the pion channel,
because its quark-meson vertex has a form
of the axial charge operator.
For evaluations of the meson exchange contributions,
it is useful to employ the $SU(4)$ spin-flavor
algebra and its Casimir values at various orders.
The many-body forces, including the Wess-Zumino vertex,
will be discussed as well.

\subsection{ $\Nc$-counting}

\begin{figure}[tb]
\vspace{0.0cm}
\begin{center}
\scalebox{0.6}[0.6] {
\hspace{-0.6cm}
  \includegraphics[scale=.38]{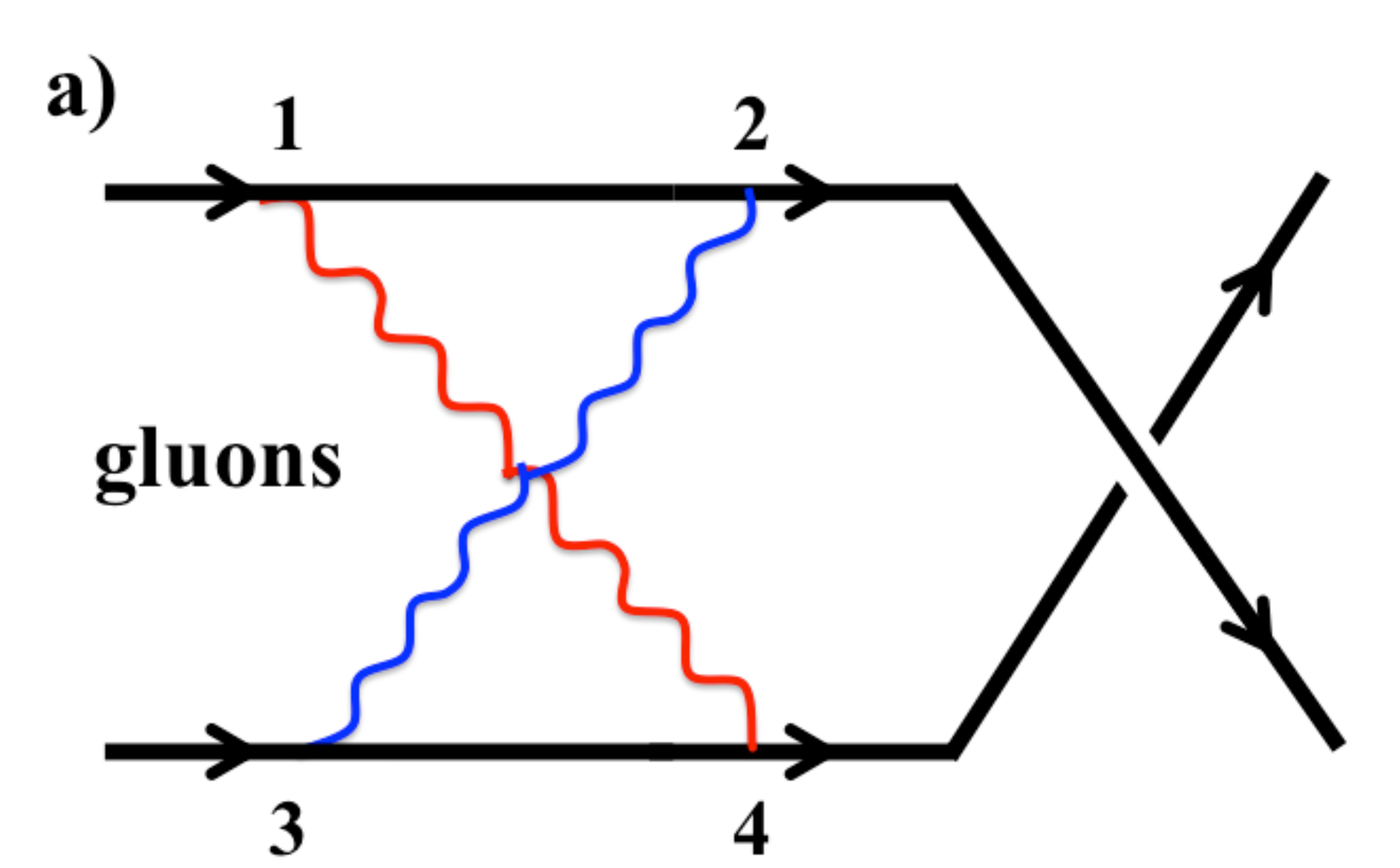} }
\scalebox{0.6}[0.6] {
\hspace{0.0cm}
  \includegraphics[scale=.38]{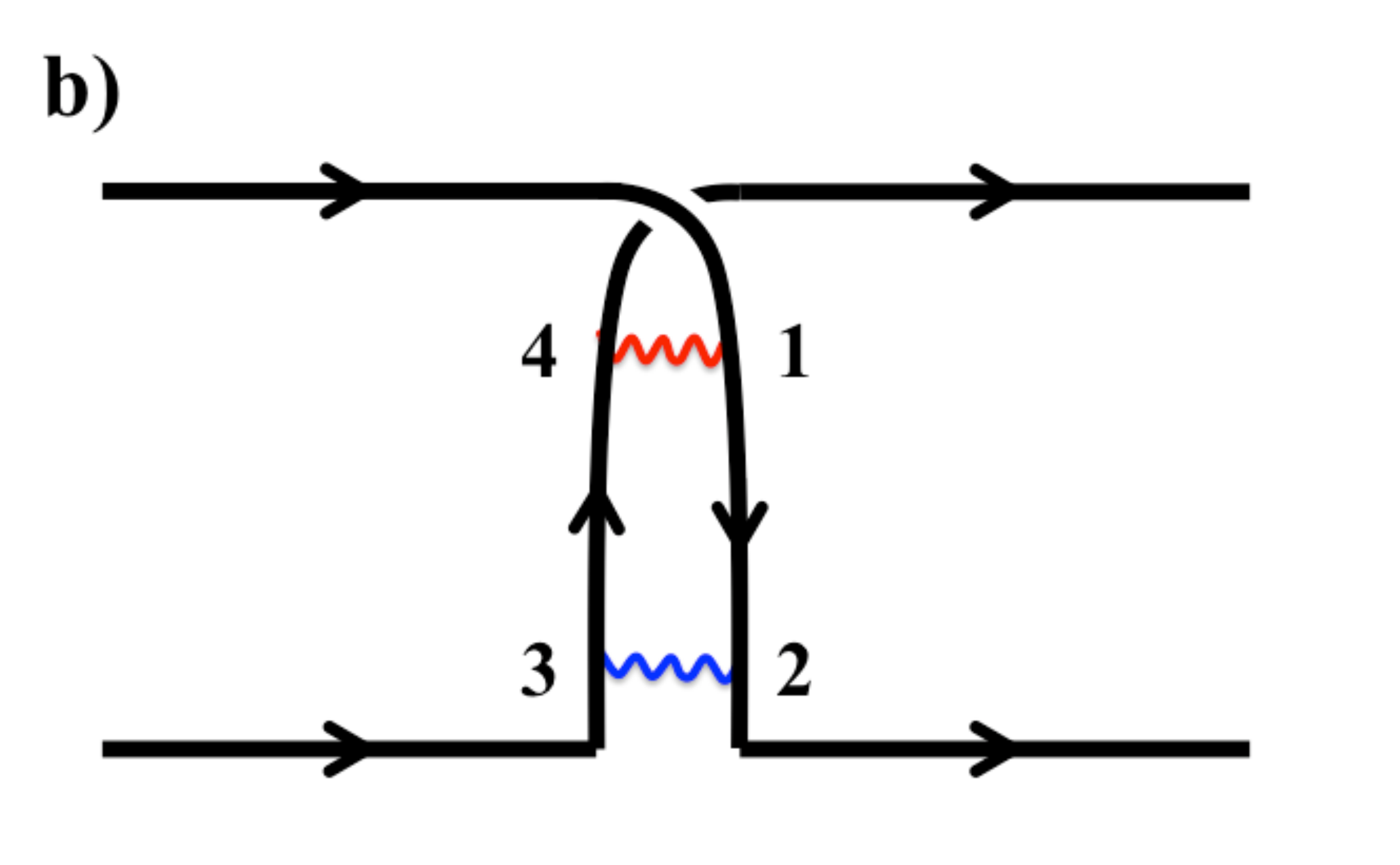} }
\end{center}
\begin{center}
\scalebox{0.6}[0.6] {
\hspace{-0.6cm}
  \includegraphics[scale=.38]{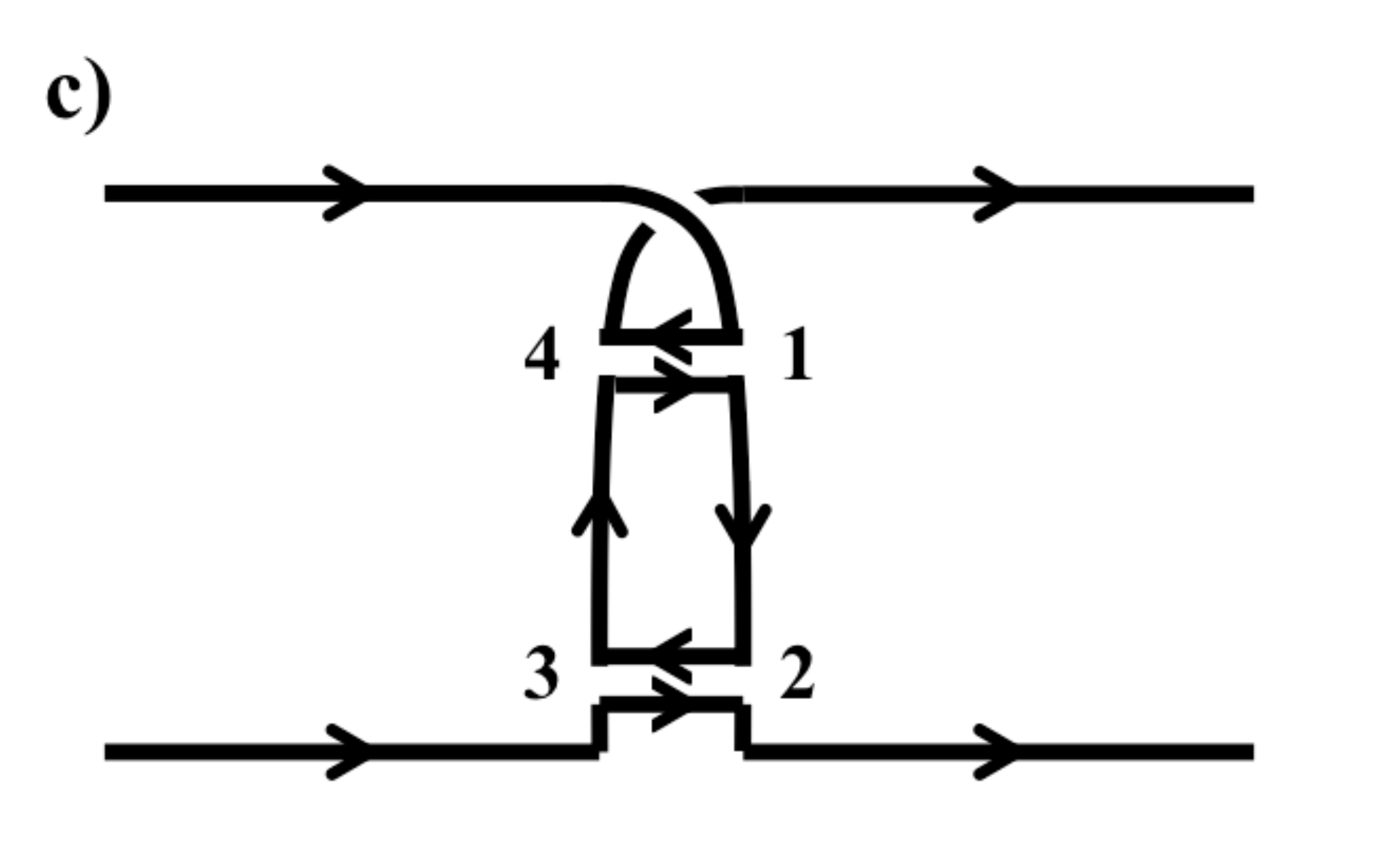} }
\scalebox{0.6}[0.6] {
\hspace{0.0cm}
  \includegraphics[scale=.38]{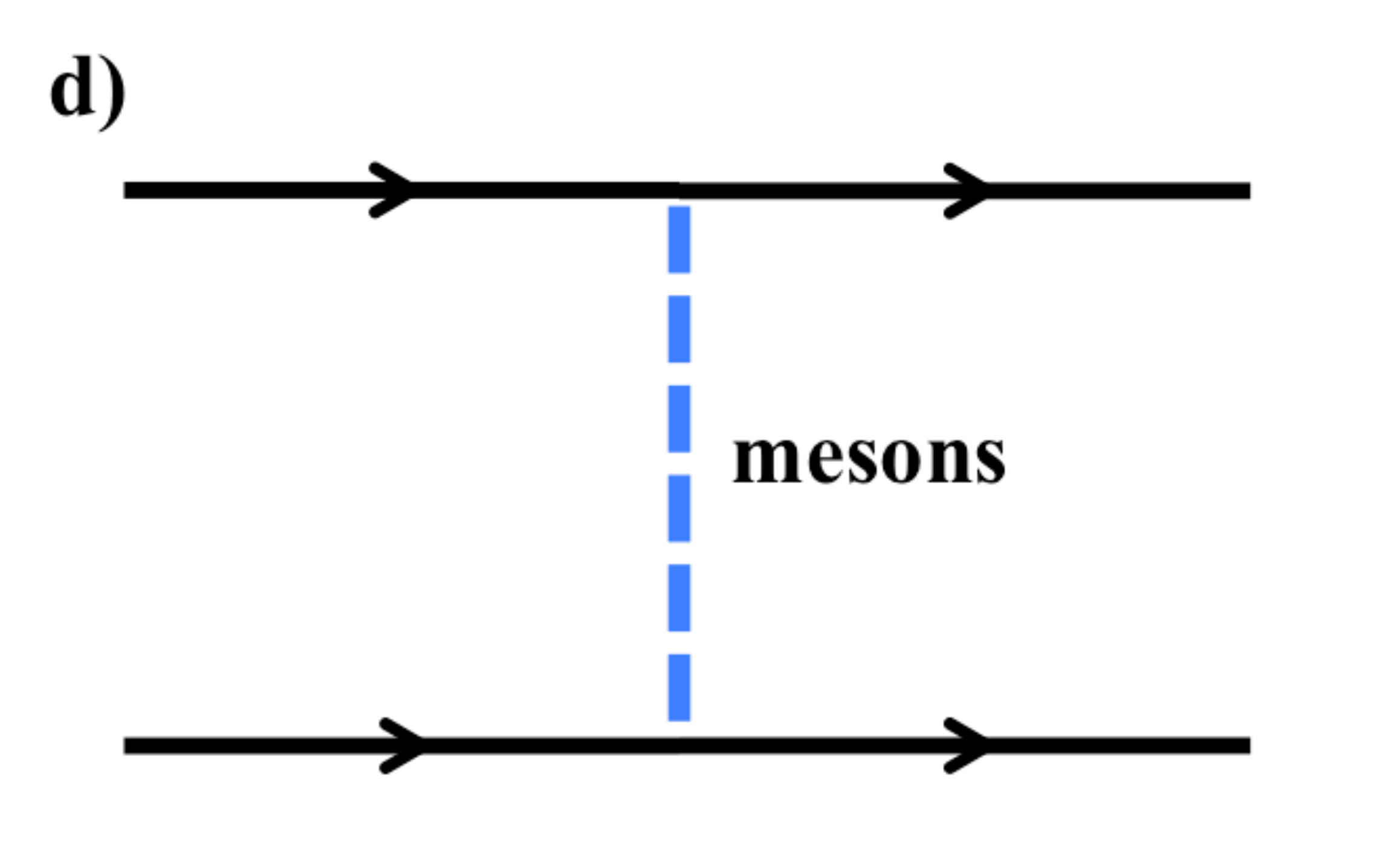} }
\end{center}
\vspace{0.0cm}
\caption{An example of two particle irreducible 
diagrams in which all gluon lines are crossed.
We attached numbers to vertices
to identify their locations after
deforming quark lines.
(a) Two gluon exchanges.
(b) A deformed version of the diagram (a)
which can be interpreted as a quark exchange.
(c) A color line representation.
There is a closed color line which provides
an enhancement factor $\Nc$.
(d) The corresponding meson exchange between quarks.
}
\vspace{0.2cm}
\label{fig:meson}
\end{figure}
Suppose diagrams in which gluons are exchanged
between two quarks.
Diagrams that are obtained by iterations
of a single gluon exchange
may be treated within the one gluon exchange potential.
On the other hand,
there are also diagrams
in which gluon exchange lines
are all crossed, and are two particle irreducible 
(Fig.\ref{fig:meson}).
For the latter, by deforming quark lines
it becomes manifest that the diagram is a planar one,
and it is natural to interpret this quark exchange diagram 
as a meson exchange between two quarks.

Here one might think that
at a distance scale among quarks inside of a baryon,
it is not appropriate to consider quark exchanges as
the meson exchanges.
Such a terminology should not be taken too literally
in this paper.
The word ``meson'' is used as
a representative of the two particle irreducible diagram
such as Fig. \ref{fig:meson}(a),
which carries the same quantum number as mesons.
The person who dislikes this terminology
can directly resum diagrams
within the language of quarks and gluons.
The real problem we cannot rigorously handle in this paper
is rather the higher towers of mesons
appearing at the distance scale $\sim \lqcd^{-1}$,
although it is likely that
such mesons do not alter the main lines of our thoughts.

\begin{figure}[tb]
\vspace{0.0cm}
\begin{center}
\scalebox{0.6}[0.6] {
\hspace{-0.6cm}
  \includegraphics[scale=.35]{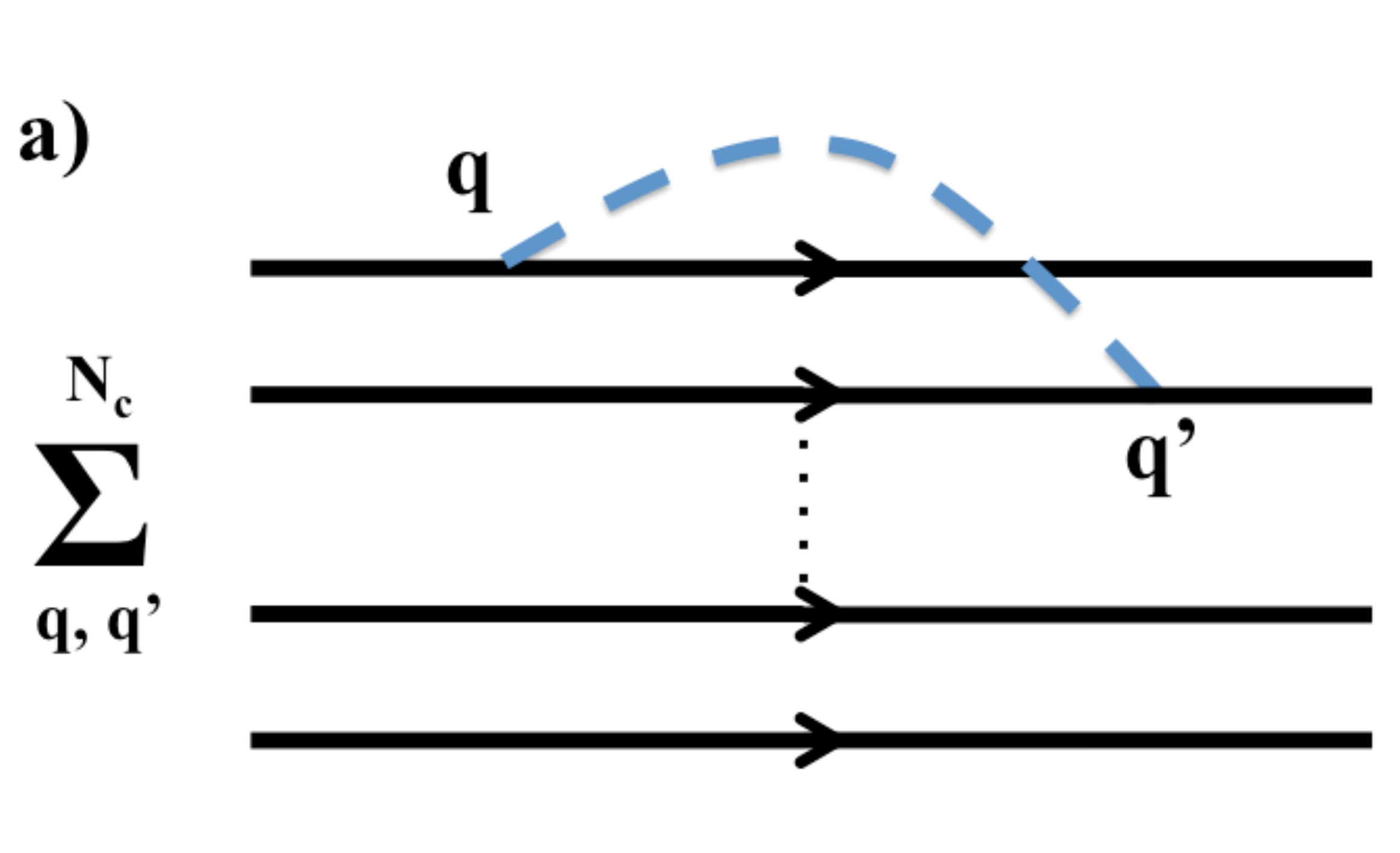} }
\scalebox{0.6}[0.6] {
\hspace{1.2cm}
  \includegraphics[scale=.35]{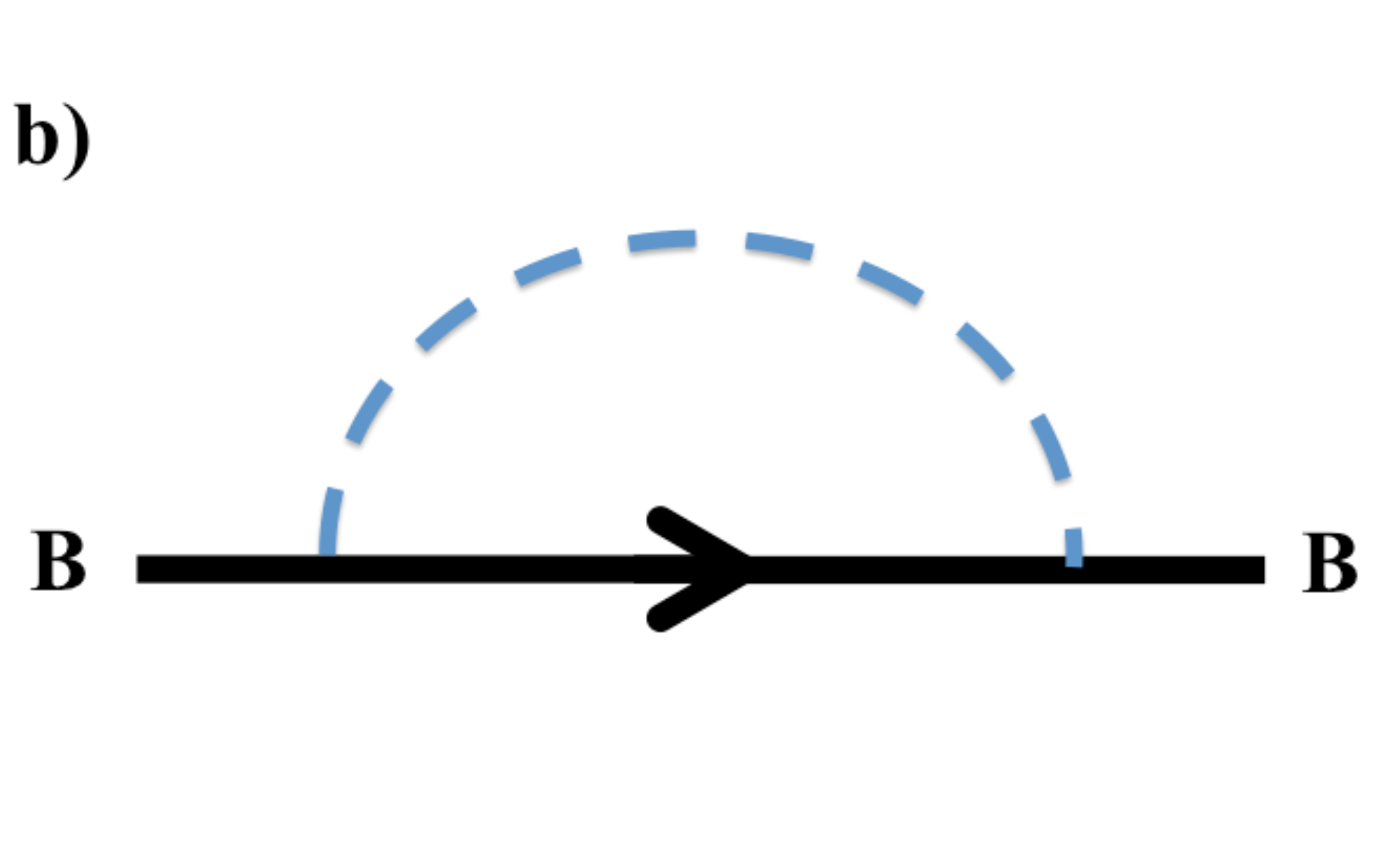} }
\end{center}
\vspace{0.0cm}
\caption{The self-energy diagram for a baryon.
(a) The meson exchange among quarks.
There are many possible ways to choose quark lines, 
$\sim \Nc^2$,
whose contributions must be summed up.
(b) The corresponding diagram 
in which a meson loop is attached to a baryon line.
}
\vspace{0.2cm}
\label{fig:baryon}
\end{figure}

The meson exchanges among quarks
can appear in the $O(\Nc)$ contributions
for the baryon self-energy,
because there are many possible ways to 
choose the quark lines.
We must keep them
in the computation of the baryon mass
as the leading order contributions.
In the language of the effective Lagrangian for mesons and baryons,
the meson exchanges among quarks
correspond to the meson loops attached to a baryon line
(Fig.\ref{fig:baryon}).

As a reminder of the self-energy evaluation, 
consider diagrams such that 
one meson is exchanged between two quarks.
Following the usual $\Nc$-counting,
the intrinsic quark-meson coupling is
given by $g_{qqM} \sim \Nc^{-1/2}$.
One can choose two quark lines
out of $\Nc$ quark lines,
so its combinatorial factor is $\sim \Nc^2$.
So if all of contributions are simply added
{\it without cancellations},
one can estimate the contributions from these processes
as $( g_{qqM} )^2 \times \Nc^2 \sim \Nc$.

Actually, whether the total of the 
meson exchange contributions
are large or not, strongly depends upon
a spin-flavor wavefunction of a baryon.

The simplest example is the charge-charge interaction
mediated by the $\omega_0$-meson exchange
(the zeroth component of the $\omega$ meson).
The $\omega_0$ couples to a quark number.
Since all quarks have quark number $+1$,
its sign always appears in the same sign.
It means that all contributions are simply additive,
so the self-energy is 
$( g_{qqM} )^2 \times \Nc^2 \sim \Nc$.
 
There are many other examples in which
various contributions cancel one another.
Consider the $\rho_0$ exchanges which are responsible
for the interactions among isospin charges.
In contrast to the quark number,
isospins of quarks may be either positive or negative, so that
their contributions can cancel one another.
For baryons with isospins of $O(1)$,
the $\rho_0$-exchange diagrams happen to 
cancel out one another,
producing negligible contributions,
$g_{qqM}^2 \times 1 \sim 1/\Nc$.
This process is important only 
for baryons with isospins of $O(\Nc)$.

\begin{figure}[tb]
\vspace{0.0cm}
\begin{center}
  \includegraphics[scale=.25]{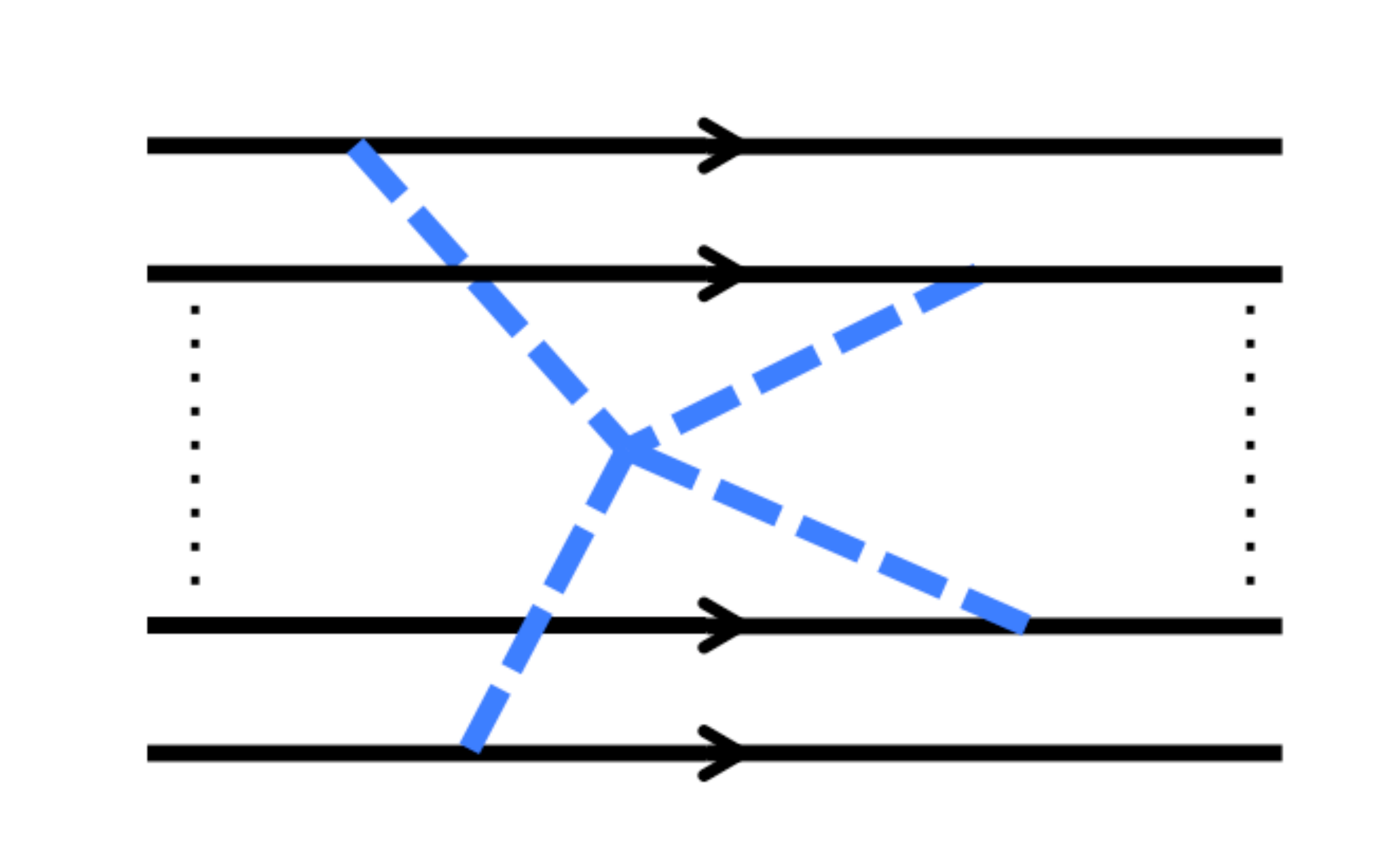} 
\end{center}
\vspace{0.0cm}
\caption{An example of the many-body forces in the baryon,
induced by mesons.
The diagram can appear at the leading $\Nc$ contribution.
}
\vspace{0.2cm}
\label{fig:multimesons}
\end{figure}

In addition to the meson exchanges between two quarks,
one must take into account many-body forces as well,
because a quark density inside of a baryon is not dilute
and there are many chances many mesons meet at once
(Fig.\ref{fig:multimesons}).
Although the $n$-meson vertex
behaves as $\sim \Nc^{1-n/2}$ and is suppressed,
there are many ways to choose $n$-quark lines
out of $\Nc$-quark lines, $\sim \Nc^n$.
Each quark-meson vertex has $\sim \Nc^{-1/2}$,
so as a total, the contributions from many-body
forces can be 
$\sim \Nc^{1-n/2} \times \Nc^n \times \Nc^{-n/2} \sim \Nc$.
In general, the tree diagrams of meson lines
can appear at the leading order of $\Nc$.

This sort of arguments are also useful 
for the classification of the two-body and many-body forces
in a baryonic matter.
In fact, it has been used to interpret
the hierarchy of the several meson exchange forces
in the nucleon-nucleon potential \cite{Kaplan:1996rk}.

\subsection{ The axial charge operator
in the $SU(4)$ spin-flavor algebra}

In this work, we will mainly focus on the
diagrams involving pions,
because we expect them most directly related to 
the axial charge of a baryon.
Considering the non-linear realization of
pions and quarks,
their couplings contain at least one derivative.
The lowest order coupling is given 
by\footnote{Our convention is
$\la 0 | 
\bar{q} \gamma_\mu \gamma_5 \frac{\tau_a}{2} q (x)
| \pi_b (q) \ra
= -\rmi q_\mu f_\pi \, \rme^{-\rmi qx}$,
$U(x)=\rme^{\rmi \pi_a \tau_a/f_\pi}$,
for which the pion decay constant is $f_\pi\simeq 93$ MeV.}
\begin{equation}
\calL_{qq\pi} =  g_A^q \, \frac{ \partial^\mu \pi_a}{2f_\pi} \, 
\bar{q} \gamma_\mu \gamma_5 \tau_a q \,.
\end{equation}
We will take the quark axial charge $g_A^q$ to be $1$
\cite{Weinberg:1990xm}.

We will take the non-relativistic approximation for quarks.
This approximation is not rigorous for the quantitative
estimates for the light quark system.
But we expect that 
the relativistic corrections change overall strengths
of the pion interactions equally for different wavefunctions,
without affecting their energy ordering.

Applying the non-relativistic approximation
for quarks in the baryons,
the above vertex is simplified to
\begin{equation}
\la B | \bar{q} \gamma_\mu \gamma_5 \tau_a q (x) | B' \ra 
~\longrightarrow~
\delta_{\mu i} \, \sum_q 
\la B| \tau^{(q)}_a \sigma^{(q)}_i \delta(\vx - \vr^{(q)} ) |B'\ra \,
\,,
\label{NR}
\end{equation}
where the zeroth component vanishes
at the non-relativistic limit,
and dominant contributions come from the spatial components.
This matrix element carries the wavefunction dependence
of the pion-baryon coupling.
We shall evaluate it using the group theory.

The axial charge operators, together with 
isospin and spin operators,
form the $SU(4)$ algebra.
Writing operators
\begin{equation}
\tau_a= \sum_q \tau_a^{(q)}\,,~
\sigma_i = \sum_q \sigma_i^{(q)}\,,~
R_{ai} = \sum_q R_{ia}^{(q)} = \sum_q \tau_a^{(q)} \sigma_i^{(q)} \,, 
\end{equation}
their commutation relations are
\begin{align}
&
[\, \tau_a, \tau_b \,] = 2 \rmi \epsilon_{abc} \tau_c \,, ~~~~
[\, \sigma_i, \sigma_j \,] = 2 \rmi \epsilon_{ijk} \sigma_k \,, 
\nonumber \\
&
[\, \tau_a, R_{bj} \,] = 2 \rmi \epsilon_{abc} R_{cj} \,, ~~~~
[\, \sigma_i, R_{bj} \,] = 2 \rmi \epsilon_{ijk} R_{bk} \,,
\nonumber \\
&
[\, \tau_a, \sigma_{j} \,] = 0 \,, ~~~~
[\, R_{ai}, R_{bj} \,] 
= 2 \rmi \epsilon_{abc} \delta_{ij} \tau_{c} 
+ 2 \rmi \epsilon_{ijk} \delta_{ab} \sigma_{k} 
\,.
\end{align}
Clearly, the operators $R_{ai}$ transform
as vectors under the operations of 
isospin or spin rotations.

In the following,
we classify the baryon states
by specifying their eigenvalues of 
the Cartan bases for the $SU(4)$ algebra.
As Cartan bases, we choose
\begin{equation}
(\, \tau_3 \,,  \sigma_3 \,,  R_{33} \,)\,,
\end{equation}
where these operators commute one another
so that we can label states with these quanta.
The states in the irreducible representation
$\calD$ can be labeled as
\begin{equation}
| \tau_3 \,, \sigma_3 \,, R_{33} \,; \calD \ra \,.
\end{equation}
For instance,
fundamental representations are formed by
single quarks
\begin{equation}
|\uup \ra = | 1,1,1\ra\,,~
|\dup \ra = | -1,1,-1\ra\,,~
|\udown \ra = | 1,-1,-1\ra\,,~
|\ddown \ra = | -1,-1,1\ra\,,~
\end{equation}
which will be building blocks
for the baryon wavefunctions.
We will give their explicit forms in the next section.

\subsection{ The processes involving pions
and their relations to the axial charge}

Now let us see how these operators appear
in the diagrams involving pions.
We classify contributions
by the order of the $R$ operators.
For instance, one pion exchange between two quarks 
is the order of $R^2$,
the diagram involving four pion emissions from four quarks
and one $4\pi$ vertex is the order $R^4$, etc. 
Later we will evaluate the expectation values of these
generators.

\subsubsection{ The $R^2$ operators: One pion exchange}
%
\begin{figure}[tb]
\hspace{0.0cm}
\begin{center}
  \includegraphics[scale=.30]{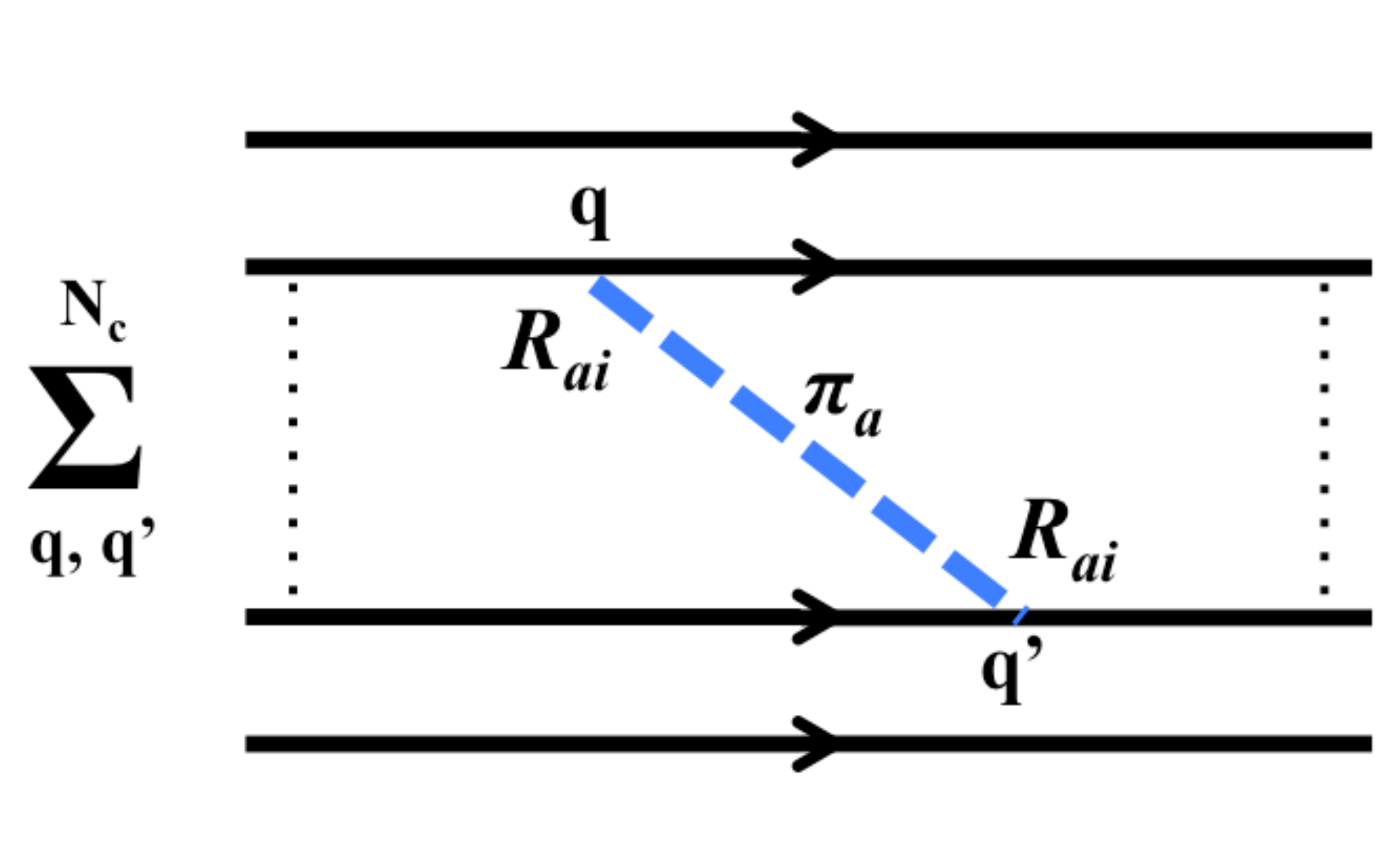} 
\end{center}
\vspace{0.0cm}
\caption{The one pion exchange contribution
to the baryon mass.
Its total size is proportional to
the baryon matrix element for the $R_{ai}^2$ operator.
}
\vspace{0.2cm}
\label{fig:R2}
\end{figure}

The simplest diagram contributing to the $R^2$ operator
is the one pion exchange among quarks.
We wish to evaluate the following matrix element
(Fig.\ref{fig:R2})
\begin{equation}
{\cal M}_{2\pi} = \la B | \,
\frac{\rmi^2}{4f_\pi^2 } \!
\int \rmd^4 x \rmd^4 y \, 
\big( \partial_i \pi_a(x) \, \bar{q} \gamma_i \gamma_5 \tau_a q (x)
\big) \, 
\big( \partial_j \pi_b(y) \, \bar{q} \gamma_j \gamma_5 \tau_b q (y) \big) 
| B \ra\,,
\label{onepi}
\end{equation}
where we dropped off the zeroth component of the
axial-vector current, taking the non-relativistic limit,
Eq.(\ref{NR}). 
${\cal M}_{2\pi}$ should be matched with the potential as
${\cal M}_{2\pi} = -\rmi \la V_{2\pi} \ra T$,
because $\la \rme^{-\rmi(H_0 +V) T } \ra$.

We shall carry out the integration of virtual lines,
leaving only quark operators.
The pion propagator is
\begin{equation}
\big\la T \, \partial_i \pi_a  (x) \, \partial_j \pi_b (y) \big\ra
= \delta_{ab} \int \! \frac{\rmd^4 p}{(2\pi)^4} \, \rme^{-\rmi p(x-y)}
\, \rmi\, 
\frac{ \,  p_i p_j \, }{\, p^2 - m_\pi^2 \,} \,.
\end{equation}
In the non-relativistic limit,
$p_0^2$ term in the denominator is 
smaller than the spatial component 
by $|\vp|/M$ where $M$ is the
constituent quark mass\footnote{This can
be seen by writing equations
in terms of the old-fashioned perturbation theory.}.
Taking the non-relativistic limit,
we decompose the propagator into
the central and tensor parts,
\begin{equation}
\frac{ \,  p_i p_j \, }{\, p^2 - m_\pi^2 \,} 
\, \longrightarrow \,
- \, \frac{\, \delta_{ij} \,}{3} \, \frac{ \vp^2 }{\, \vp^2 + m_\pi^2 \,} 
- \left( \frac{ p_i p_j }{ \vp^2} -  \frac{\, \delta_{ij} \,}{3} \right)
\frac{ \vp^2 }{\, \vp^2 + m_\pi^2 \,} \,.
\end{equation}
The second term is activated
in the intermediate processes such that
two quarks in the $S$-wave are scattered into the $D$-wave.
We shall ignore such a contribution,
and will take into account only the first term.
Together with these simplifications,
\begin{equation}
\big\la T \, \partial_i \pi_a  (x) \, \partial_j \pi_b (y) \big\ra
\, \rightarrow \,
-\, \frac{ \rmi \delta_{ab} \delta_{ij} }{ 3 }
\, \delta(x_0-y_0) \!
\left(
\delta^3(\vx-\vy)
- \frac{m_\pi^2}{4\pi} \frac{\, \rme^{-m_\pi |\vx - \vy|} \,}{|\vx-\vy| } 
\right) .
\end{equation}
The reason why the $\delta$-function appears
is that the derivative coupling of pions to quarks
becomes larger for larger momentum transfer.
It cancels out the $1/\vp^2$ behavior of the pion propagator,
so that the pion propagations with relatively high momenta
are not suppressed (although in a more realistic treatment
there should be the UV cutoff due to form factors).
On the other hand, the second term
is proportional to the pion mass,
so disappears in the chiral limit.

In the chiral limit together with aforementioned approximations,
Eq.(\ref{onepi}) is reduced to
\begin{align}
& {\rm Eq}.(\ref{onepi}) \rightarrow
\frac{\rmi}{12 f_\pi^2 } 
\int \rmd^4x 
\, \la B | 
\bar{q} \gamma_i \gamma_5 \tau_a q (x) \,
\bar{q} \gamma_i \gamma_5 \tau_a q (x)
| B \ra \, 
\nonumber \\
& \hspace{-0.0cm}
= \frac{\rmi T}{12 f_\pi^2 } 
\int \rmd^3 \vx \, 
\sum_{q,q'} \sum_{B'}  \,
\la B| R_{ai}^{(q)} \delta(\vx - \vr^{(q)} ) |B'\ra
\la B'| R_{ai}^{(q' )}\delta(\vx - \vr^{(q')} ) |B\ra
\,. 
\label{T2}
\end{align}
We mainly consider all quarks in $|B\ra$ occupy the
same spatial orbit, $\varphi$.
To have nonzero matrix elements,
quarks in $|B'\ra$ must occupy $\varphi$
except the $q$-th quark which hits the $\delta$-function.
We ignore $|B'\ra$ including a radial excitation of a quark,
and consider only $|B'\ra$ with all quarks in the spatial orbit $\varphi$.
Then the equation is further simplified by
summing the intermediate state $|B'\ra$,
\begin{equation}
{\rm Eq}.(\ref{T2})
\simeq 
\frac{\rmi T}{12 f_\pi^2 } 
\int \rmd^3 \vx \, 
\left| \varphi (\vx) \right|^4 \,
\la B|R_{ai}^2 |B\ra_{ {\rm SF}} 
\,.
\end{equation}
where $|B\ra_{{\rm SF}}$ means the spin-flavor part of $|B\ra$.
Therefore the potential is
\begin{equation}
\la V_{2\pi} \ra 
\simeq - \, \frac{1}{\, 12 f_\pi^2 \,} 
\int \rmd^3 \vx \,\, 
| \varphi (\vx) |^4 \,
\la B|R_{ai}^2 |B\ra_{ {\rm SF}} 
\sim
- \, \frac{\, \lqcd^3\, }{12 f_\pi^2 } \,
\la B|R_{ai}^2 |B\ra_{ {\rm SF}} \,,
\end{equation}
which reduces the baryon mass.
Here we assumed that the spatial extent of $\varphi(\vx)$
is $\sim \lqcd^{-1}$,
so that $|\varphi(\vx)|^2 \sim \lqcd^{3}$ from the normalization.

\subsubsection{ The $R^3$ operators: Wess-Zumino vertex}
%
\begin{figure}[tb]
\vspace{0.0cm}
\begin{center}
\hspace{-0.8cm}
  \includegraphics[scale=.30]{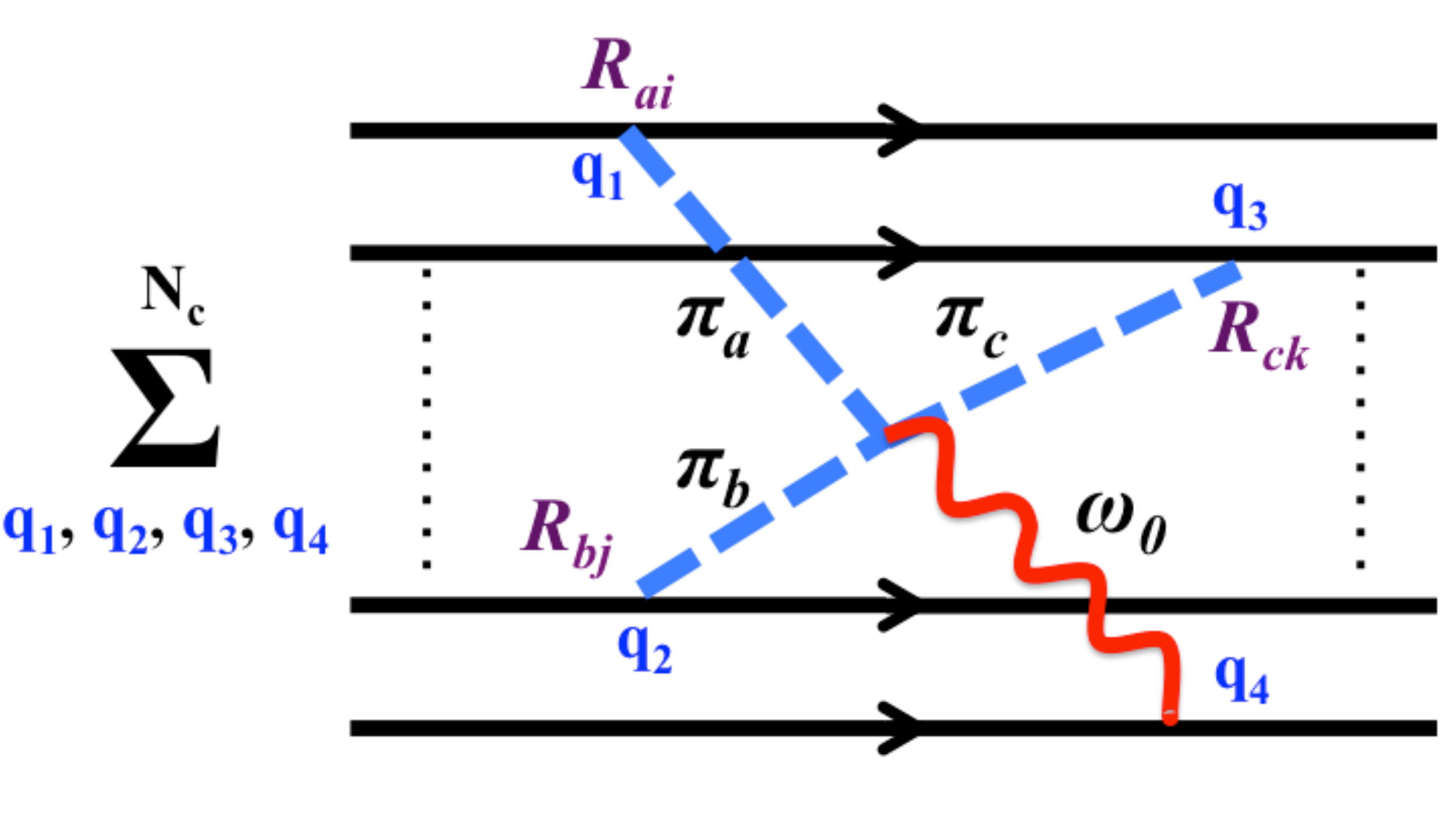} 
\end{center}
\vspace{0.0cm}
\caption{The $\omega$-$3\pi$ contribution
to the baryon mass, mediated by the 
Wess-Zumino vertex.
Its total size is proportional to
the baryon matrix element for the 
$\epsilon_{abc}\epsilon_{ijk} R_{ai}  R_{bj}  R_{ck}$ operator.
}
\vspace{0.2cm}
\label{fig:WZ}
\end{figure}
Similar manipulations can be done also for
higher order products of the $R$ operators.
Here of particular interest is
the $\omega$-$3\pi$ vertex coming from
the Wess-Zumino term (Fig.\ref{fig:WZ}).
The diagrams involving the $\omega$ meson
are very important
because $\omega_0$ strongly couples to 
a baryon
due to its large quark number density of $O(\Nc)$.
In fact, in the Skyrme or chiral soliton models,
this term is responsible for 
a repulsive force of $O(\Nc)$ between
a quark number density and a topological density,
which stabilizes the configuration of the chiral soliton.

The quark-$\omega$ meson coupling appears in the form
\begin{equation}
\calL_\omega
= 
- g_\omega \omega_\mu \, \bar{q} \gamma^\mu q\,.
\end{equation}
If we take the sign of $g_\omega$ in this way,
the Wess-Zumino term for the $\omega-3\pi$ coupling
is given by\footnote{The expression can be intuitively checked
from the $\omega$-quark vector coupling,
$\calL_{qq\omega} = - g_\omega \omega_\mu J^\mu$,
with the Goldstone-Wilczek's topological 
quark number current, 
\begin{equation}
J^\mu = - \frac{\Nc}{24\pi^2} \,
\epsilon^{\mu \nu \alpha \beta} \,
\tr_f\left[ 
(U^\dag \partial_\nu U ) 
( U^\dag \partial_\alpha U )
( U^\dag \partial_\beta U ) 
\right] \,,
\end{equation}
where the trace is taken over
the $SU(2)$ fundamental representations. 
}
\begin{equation}
\calL_{\omega-3\pi} 
= \frac{ \Nc g_\omega}{12\pi^2 f_\pi^3} \, 
\epsilon^{\mu \nu \alpha \beta}
\epsilon_{abc} \, \omega_\mu 
\, (\partial_\nu \pi_a) \, (\partial_\alpha \pi_b) \, (\partial_\beta \pi_c) \,.
\end{equation}
When evaluating this contribution for the baryon states,
the term involving $\omega_0$ field
is much larger than terms with $\omega_i$'s.
Thus we consider only
\begin{equation}
\calL_{\omega_0-3\pi} 
= - \frac{ \Nc g_\omega}{12\pi^2 f_\pi^3} \, 
\epsilon_{ i j k}
\epsilon_{abc} \, \omega_0 
\, (\partial_i \pi_a) \, (\partial_j \pi_b) \, (\partial_k \pi_c) \,.
\end{equation}
As before, we draw a diagram and integrate
the virtual meson propagations,
leaving the matrix element for quark operators.
We compute
\begin{align}
\hspace{-0.5cm}
{\cal M}_{\omega-3\pi}
&= 
\rmi^5 \left( \frac{-1}{2 f_\pi} \right)^{\!3} ( -g_\omega)
\times
\frac{ - \Nc g_\omega }{12\pi^2 f_\pi^3} 
\int \rmd^4x\, \rmd^4y\, \rmd^4z\, \rmd^4w\, \rmd^4u \,
\la B | \Gamma_{\omega-3\pi} |B \ra \nonumber \\
&=
- \rmi \, \frac{\, \Nc  g_\omega^2 \, }{96\pi^2 f_\pi^6}
\int \rmd^4x\, \rmd^4y\, \rmd^4z\, \rmd^4w\, \rmd^4u \,
\la B | \Gamma_{\omega-3\pi} |B \ra \,,
\end{align}
where
\begin{align}
\Gamma_{\omega-3\pi} 
& =
\big( \bar{q} \gamma_i \gamma_5 \tau_a q (x) \big)
\big( \bar{q} \gamma_j \gamma_5 \tau_b q (y) \big)
\big( \bar{q} \gamma_k \gamma_5 \tau_c q (z) \big)
\big( \bar{q} \gamma_0 q (w) \big)
\nonumber \\
& \times 
\big( \partial_i \pi_a(x) \big) \big( \partial_j \pi_b(y) \big) 
\big( \partial_k \pi_c(z) \big) \, \omega_0(w)
\nonumber \\
& \times
\epsilon_{lmn}
\epsilon_{def} \, \omega_0 (u) \, 
\big(\partial_l \pi_d (u) \big) 
\big(\partial_m \pi_e (u) \big)
\big(\partial_n \pi_f (u) \big) \,.
\end{align}
The propagator of the $\omega$ meson,
in the non-relativistic approximation, is
\begin{align}
\big\la T \omega_0 (w) \, \omega_\mu (u) \big\ra
&= \int \! \frac{\rmd^4 p}{(2\pi)^4} \, \rme^{-\rmi p(w-u)}
\frac{ -\rmi}{p^2-m_\omega^2} 
\left( g_{0 \mu} - \frac{p_0 p_\mu}{m_\omega^2} \right)
\nonumber \\
&\simeq 
\rmi g_{0\mu} \, \delta(w_0-u_0) \, G_\omega( \vw - \vu)\,.
\end{align}
where 
\begin{equation}
G_\omega( \vw - \vu)
= \frac{\rme^{-m_\omega |\vw -\vu|} }{\, 4\pi |\vw - \vu| \, } \,.
\end{equation}
Assembling these approximations together,
our vertex now takes the form
%
%
\begin{equation}
{\cal M}_{\omega-3\pi}
\rightarrow
 \rmi \, \frac{\, \Nc g_\omega^2 T\, }{2^4 3^3 \pi^2 f_\pi^6}
\int \rmd \vx\, \rmd  \vw \,
\la B | R^{(3)} (\vx) \, G_\omega(\vx-\vw) \, Q_0(\vw) | B \ra \,,
\label{3piwz}
\end{equation}
where we defined the cubic product of the
axial charge operator,
\begin{equation}
R^{(3)} (\vx) \equiv 
\epsilon_{abc} \epsilon_{ijk} \,
R_{ai} R_{bj} R_{ck} (\vx) 
\,, ~~~R_{ai} (\vx) = \sum_q R_{ai}^{(q)}  \delta(\vx - \vr^{(q)} ) \,, 
\end{equation}
and the quark number operator is
$Q_0 (\vw) = \bar{q} \gamma_0 q (\vw)$.
The quark number is very large, $\Nc$, 
for any baryon states.
Again assuming that
all quarks occupy the spatial orbit $\varphi$,
then Eq.(\ref{3piwz}) becomes
\begin{equation}
\rmi \frac{\,\Nc g_\omega^2 T\, }{\, 2^4 3^3 \pi^2 f_\pi^6\, }
\, \la B | R^{(3)} | B \ra_{ {\rm SF} } 
\int \rmd \vx\, \rmd  \vw \,\, |\varphi(\vx)|^6 \,
G_\omega(\vx-\vw) \times 
\left( \Nc  \, |\varphi(\vw)|^2 \right)\,.
\end{equation}
This takes the form of the interaction
between the distribution of the $R^3$ operator
and the quark number density. 
Finally the potential energy is
\begin{equation}
\la V_{\omega-3\pi} \ra
= - \frac{\, \Nc^2 g_\omega^2 \, }{\,2^4 3^3\pi^2 f_\pi^6\,}
\, \la B | R^{(3)} | B \ra_{ {\rm SF} } 
\int \rmd \vx\, \rmd  \vw \,\,
 |\varphi(\vx)|^6 \,
G_\omega(\vx-\vw) \, |\varphi(\vw)|^2 \,.
\end{equation}
As we will see,
the $R^3$ term appears to be negative and $O(\Nc^3)$.
The overall factor is $\Nc^2 g_\omega^2/f_\pi^6 \sim 1/\Nc^2$,
so $V_{\omega-3\pi}$ gives the 
energetic cost of $O(\Nc)$.

\section{Hedgehog, conventional, and dichotomous 
nucleon wavefunctions
and their axial charges}
\label{Wavefunctions}

In this section,
we show explicit forms of the hedgehog,
conventional, and dichotomous wavefunctions.
Then we compute 
the expectation value of the $R_{33}$ operator.

\subsection{A hedgehog nucleon wavefunction}

The baryon wavefunction giving a hedgehog type
configuration has zero grandspin,
$\vec{G} \equiv \vec{\tau} + \vec{\sigma} =0$. 
The basic building block is a single 
quark wavefunction
\begin{equation}
\left( \frac{ u\! \down - d \!\up }{\sqrt{2}} \right) \,,
\end{equation}
which vanishes under the operation 
$\vec{G} \equiv \vec{\tau} + \vec{\sigma}$. 
(This choice is unique.
Other wavefunction such as 
$\left( u\! \down + d \!\up \right) $
does not vanish under the operation of $\tau_1 + \sigma_1$, etc.)
The hedgehog baryon wavefunction is
\begin{equation}
| H \ra
=  \left( \frac{ u\! \down - d\! \up }{\sqrt{2} } \right)^{\Nc} \,.
\end{equation}
This is just a product of the same single quark wavefunction,
so it is maximally symmetric in spin-flavor.
The special property is that
$|H\ra$ has a definite $R_{33}$ eigenvalue.
Indeed, in the $SU(4)$ bases ($|\tau_3, \sigma_3, R_{33}\ra$)
one can write
\begin{equation}
\left( u\! \down - d \!\up \right)
= | 1, -1, -1 \ra - | -1, 1, -1\ra \,,
\end{equation}
where the eigenvalue $R_{33}$ is common, so that
its any products have a definite $R_{33}$ eigenvalue.
But they are a linear superposition of the
different $(\tau_3, \sigma_3)$.
In these bases, the hedgehog wavefunction is written as
\begin{equation}
|H \ra 
= \left( \frac{1}{\sqrt{2} } \right)^{\!\Nc}
\sum_{n= 0}^{\Nc} \, \sqrt{ {}_{\Nc} C_n}\, 
(-1)^{n} \, | \Nc-2n, -(\Nc-2n), -\Nc \ra_{ {\rm MS}} \,,
\end{equation}
where $|\cdots \ra_{ {\rm MS} }$ means that the state is 
maximally symmetrized.
Here a factor $\sqrt{ {}_{\Nc} C_n}$ is included 
to normalize the state as 
$\la \tau_3, \sigma_3,R_{33} | \tau_3, \sigma_3,R_{33} \ra_\rMS
=1$\footnote{For 
instance, 
\begin{align}
\left( \udown - \dup \right)^{2}
&
= (\udown)^2 + (\dup)^2 
- \left( \udown \dup + \dup \udown \right)
\nonumber \\
&
= |2,-2,-2\ra_\rMS + |-2,2,-2\ra_\rMS
- \sqrt{2} \, |0,0,-2\ra_\rMS \,.
\end{align}
}.

All the states contained in the hedgehog wavefunction
have large $R_{33}$ eigenvalues of $O(\Nc)$.
So fluctuations of the $R_{33}$ charge, that are caused by
the $SU(4)$ breaking terms in the Hamiltonian
and happen in the time evolution,
can be regarded as small fluctuations
around the big mean value of $O(\Nc)$.
Thus the axial charge of the hedgehog state
may be treated as classical and static mean field.
We will use this observation
when we give a non-perturbative consideration
for the baryon self-energy.

The axial charge of the hedgehog state
can be computed readily,
and is given by $\la H| R_{33} | H \ra = -\Nc$,
hence $g_A^{ {\rm H} } = -\Nc$.
The minimal eigenvalue $R_{33}$ for $\Nc$ quarks
are $-\Nc$, so the hedgehog state saturates
the lower bound of the baryon axial charge $g^B_A$.

\subsection{A conventional nucleon wavefunction}

The conventional nucleon wavefunctions
have definite isospins and spins,
while they are superpositions of different axial charge states.
To construct states with small isospins and spins,
it is convenient to prepare a
$|I=0,S=0\ra$ pair wavefunction as a building block:
\begin{align}
|D\ra 
&= \frac{1}{2} 
\left| u d - du \right\ra 
\otimes
\left| \up \down - \down \up \right\ra
\nonumber \\
&= 
\frac{1}{2} 
\left| u \!\up d\! \down + \,  d\! \down u\! \up \right\ra
- 
\frac{1}{2} 
\left| u \!\down d\! \up + \,  d\! \up u\! \down \right\ra
\nonumber \\
&= 
 \frac{1}{\sqrt{2} }  \big(
| 0,0, 2 \ra_{{\rm MS}} - | 0,0, -2 \ra_{{\rm MS}} 
\big) \,,
\end{align}
which is a superposition of pair wavefunctions with
different axial charge states, $R_{33} = \pm 2$.

Before going further,
we stress that
the construction of a nucleon from diquarks
has nothing to do with any dynamical assumptions
such as diquark correlations.
In fact, at large $\Nc$ the diquark correlation is
suppressed.
Here we simply use the diquarks just for
the convenience to
prepare appropriate spin-flavor quantum numbers.
Needless to say,
we can arrive at the same wavefunction
without using the diquark pair wavefunctions.

By just taking a direct product of these states,
we can obtain spin-flavor functions
with $I=S=0$.
To construct $\Nc$ odd nucleons,
we must further add one more quark
to achieve the color singletness and
appropriate spins.
Then we totally symmetrize the spin-flavor-space 
wavefunctions for products
of diquarks and an unpaired quark.
Usually a nucleon with a maximally symmetric
spin-flavor wavefunction
are supposed to be the lowest energy state,
because in this way one can place all of $\Nc$ quarks
in the $S$-wave spatial orbit
relative to the center of the nucleon.

Let us consider $|p^{ {\rm c} } \!\up\ra$,
conventional proton wavefunction, as an example.
Its spin-flavor wavefunction is 
($\calS$ is the symmetrization operator)
\begin{align}
|p^{ {\rm c} } \!\up \ra_{ {\rm SF} } 
&= \calN \calS \big[ |D\ra^{\nd} \otimes |\uup\ra \big] 
= \calN 
 \calS 
\left[ \big( 
\left| u \!\up d\! \down \right\ra
-  
\left| u \!\down d\! \up \right\ra
\big)^{n_d} \otimes | \uup\ra \right]
\nonumber \\
&= \calN \sum_{n=0}^{n_d}
\, {}_{n_d} C_n\, (-1)^n \, \calS 
\big[\, |\uup\ra^{n+1} \, |\ddown\ra^n \,
| \udown \ra^{\nd-n} \, |\dup\ra^{\nd-n} \, \big] \,,
\end{align}
where $\Nc = 2\nd +1 $
and $\calN$ is the normalization factor.
For the $SU(4)$ bases, it is written as
a superposition of states with different $R_{33}$ eigenvalues.
Using the $R_{33}$ eigenvalues for 
$(\uup, \ddown, \udown, \dup)=(1,1,-1,-1)$,
we can write
\begin{equation}
|p^{ {\rm c} }\! \up\ra_{ {\rm SF} }
= \calN' \sum_{n=0}^{n_d} \, {}_{n_d} C_n\, (-1)^n \,
|\, 1\,, 1\,, 4n +1 -2\nd \,\ra_{ {\rm MS} } \,.
\end{equation}
This spin-flavor wavefunction
contains states with both positive 
and negative $R_{33}$ eigenvalues.
The mean value and fluctuations of the $R_{33}$ charge
can be comparable during the time evolution.
In addition,
this makes the evaluation of $R_{33}$ expectation value
for the conventional nucleon state
much more nontrivial than for the hedgehog state.
In fact, 
the contributions from states with positive and
negative $R_{33}$ eigenvalues might 
strongly cancel one another.
Our proposal of the dichotomous nucleon wavefunction,
which will be discussed next,
was motivated by this observation.

The axial charge of the conventional nucleon wavefunction
has been computed in seminal works,
and is given by
\begin{equation}
g_A^{ {\rm c} } 
= \la p^{ {\rm c} } \!\up \!| R_{33} | \, p^{ {\rm c} } \! \up\ra_{ {\rm SF} }
= | \calN' |^2
\sum_{n=0}^{n_d} \, \left( {}_{n_d} C_n \right)^2 \,
(4n +1 -2\nd )
= \frac{\, \Nc+2 \,}{3} \,,
\end{equation}
where the contribution of $O(\Nc)$ survives
even after cancellations.
Flipping either spin or isospin
changes the sign of the axial charge,
so $g_A^{ {\rm c} }$ can be $\pm (\Nc+2)/3$ for
nucleons.

\subsection{A dichotomous nucleon wavefunction}

The dichotomous wavefunction
is constructed in such a way
to make a nucleon axial charge $g_A$ small.
To see how to achieve this,
let us first consider 
the maximally symmetrized spin-flavor wavefunction
made of $\nd=(\Nc-1)/2$ diquarks with $I=S=0$,
\begin{equation}
\calS \big[ |D\ra^{\nd} \big] 
=
\sum_{n=0}^{n_d} \, {}_{n_d} C_n\, (-1)^n \,
|\, 0\,, 0\,, 4n  -2\nd \,\ra_{ {\rm MS} } \,.
\end{equation}
If we take the expectation value of $R_{33}$ for
this state,
we have
\begin{equation}
\calS \big[ \la D |^{\nd} \big] \, R_{33} \,
\calS \big[ |D\ra^{\nd} \big] 
=
\sum_{n=0}^{n_d} \, \left( {}_{n_d} C_n \right)^2 \,
(4n  -2\nd )
= 0 \,,
\end{equation}
where the contributions to the axial charge
completely cancel out.
From this,
we can see that
in the conventional nucleon,
the $O(\Nc)$ contribution arises
from the overlap between the diquark part
and an unpaired quark.

Therefore, if we place the unpaired quark
in a spatial orbit different from
other $(\Nc-1)/2$ quarks,
the $O(\Nc)$ contribution to the axial charge disappears.
The dichotomous wavefunction for a proton with spin up 
is written as
\begin{equation}
|\, p^{ {\rm d} } \! \up \ra
= \calN \, \calS \big[ |D; A \ra^{\nd} \otimes | \uup;B \ra\big] \,,
\end{equation}
where the symmetrization is done for
spin-flavor-space wavefunctions,
by definition.
Here $A$ and $B$ characterize spatial wavefunctions
of quarks participating in the diquarks
and an unpaired quark, respectively.
When $A$ and $B$ are taken to be completely orthogonal,
the axial charge is saturated by that of the unpaired quark,
\begin{equation}
g_A^{ {\rm d} } 
= \la \, p^{ {\rm d} } \! \up |\, R_{33} \,
|\, p^{ {\rm d} } \! \up \ra
= \la \uup|\, R_{33} \, |\, \uup\ra = 1\,.
\end{equation}
So for the dichotomous nucleons,
$g_A^{ {\rm d} } = \pm 1$.
The more general case where $A$ and $B$ are
not completely orthogonal 
was computed in Ref.\cite{Hidaka:2010ph}.

To avoid confusions,
it is worth emphasizing that
the dichotomous wavefunction
is {\it not} an irreducible representation 
of the $SU(4)$ symmetry.
We label quark bases with
the eigenvalues of the $SU(4)$ generators,
just for a matter of convenience to organize
the computations of quantities related to
the axial charge operators.
Thus the use of the $SU(4)$ bases is not mandatory step.
In fact, the states built from them need not be
the irreducible representations of the $SU(4)$ symmetry
because the Hamiltonian we are using
is not $SU(4)$ symmetric.
The point is that we are trying to 
distinguish different nucleon wavefunctions
by the $SU(4)$ violating contributions.
Whether the $SU(4)$ symmetry turns out to be good or not
can be determined only after 
solving the dynamical problems.

The main problem when we try to regard the dichotomous
state as the ground state is 
the kinetic energy cost when placing an unpaired quark
in an orbit different from the others.
Its energetic cost is $O(\lqcd)$.
The self-energy contributions from pions
must reduce the nucleon energy,
compensating this energy cost.

\section{Perturbative analyses}
\label{Pert}

Now we have prepared all ingredients
necessary for our perturbative considerations.
In this section,
we compare the nucleon mass
for different wavefunctions, 
at the level of the perturbative contributions 
up to $O(R^3)$ operators.
It turns out that the hedgehog, conventional,
and dichotomous wavefunctions
give the same self-energy at the leading $\Nc$ 
contribution.
At NLO of $\Nc$, the signs flip
for different orders of the $R$-operators,
so we cannot conclude which wavefunctions
give the lowest energy at this level of analyses.

We will also argue specialities of the hedgehog state
in these perturbative estimates,
which will justifies the static mean field picture
of the pion fields for the hedgehog quark wavefunction.

\subsection{The contributions from the $R^2$ operator}

To compute the one pion exchange contributions
proportional to $R_{ai}^2$,
we use the relation
\begin{equation}
R_{ai}^2 = C_2 - \tau_a^2 - \sigma_i^2 \,.
\label{R^2}
\end{equation}
where $C_2$ is the second order Casimir.
The largest contribution to $R_{ai}^2$ comes from $C_2$.
For the maximally symmetric spin-flavor wavefunction
of $n$ quarks,
$C_2$ is given by
\begin{equation}
C_2 \big( {\rm MS}(n) \big)
= 3 n ( n+4 )\,,
\end{equation}
(see Appendix. \ref{Casimir2}).
The hedgehog and conventional nucleon wavefunctions
belong to the maximally symmetric spin-flavor representation
of $\Nc$ quarks,
and the expectation value of $C_2$ 
is given by $3\Nc(\Nc+4)$.

On the other hand,
in the dichotomous wavefunction,
$(\Nc-1)$ quarks and an unpaired quark
have no overlap so that
the expectation value of the Casimir
is given by a sum of that of the 
$(\Nc-1)$ quarks and one unpaired 
quark\footnote{To see this calculation explicitly,
the use of Eq.(\ref{tosee}) may be helpful.}.
Then $C_2$ is given by 
$3(\Nc-1)(\Nc+3) + 15 =3\Nc^2+6\Nc +6 $,
which differs from other two wavefunctions
by $O(\Nc)$.

The expectation values of $\tau_a^2$ and $\sigma_i^2$
are trivial for the conventional and dichotomous 
wavefunctions with definite isospin and spin quantum numbers.
The values are $\la \tau_a^2 \ra = \la \sigma_i^2 \ra=3$.

On the other hand, for the hedgehog state
as a mixture of various isospin and spin states,
we need a little manipulations.
The hedgehog state contains
isospins and spins of $O(\Nc)$,
so naively one might expect
$O(\Nc^2)$ contributions for $\la \tau_a^2 \ra$
and $\la \sigma_i^2 \ra$.
It turns out, however, that 
$\la \tau_a^2 \ra = \la \sigma_i^2 \ra =3\Nc$
due to cancellations of $O(\Nc^2)$ contributions
(see Appendix. \ref{some}).

Assembling all these results,
the one pion exchange contribution is proportional to
\begin{equation}
- \, \la R_{ai}^2 \ra 
= - 3\Nc^2 - 6 \times
\left\{
\begin{matrix}
~ \Nc ~~~& ( {\rm hedgehog} )\\
~ 2\Nc -1 ~~~& ~~~~({\rm conventional})\\
~ \Nc  ~~~& ~~~~({\rm dichotomous})
\end{matrix}
\right.
\label{R2Nc}
\end{equation}
after multiplying the overall sign for the potential energy.
We do not find the difference at
the leading order of $\Nc$.
The conventional wavefunction acquires
the largest attractive force at $O(R^2)$.
Since the intrinsic quark-meson coupling is 
$g_{\pi qq} \sim \Nc^{-1/2}$,
the energy difference appears to be
$\sim g_{\pi qq}^2 \times \Nc \sim \lqcd$.

Through Eq.(\ref{R^2}),
one can observe several interesting consequences
for the excited states in the maximally symmetric representation.
First, the single pion exchanges 
produce the same mass splitting rule for
the higher isospin and spin multiplets.
Second, for baryons with larger the isospins and/or spins,
the self-energy contribution from the pion exchange
is less important.

Finally we have to mention about the size of the
$1/\Nc$ corrections.
The expression $(\ref{R2Nc})$ 
explicitly contains
the subleading terms of $1/\Nc$,
and one can observe that the $1/\Nc$ corrections
are substantial for $\la R_{ai}^2 \ra$.
We will return to this issue at Sec. \ref{Discussion}.

\subsection{The contributions from the $R^3$ operator}

To compute the $R^3$-operators,
we will use the relation,
\begin{equation}
\epsilon_{ijk} \epsilon_{abc} R_{ai} R_{bj} R_{ck}
= 
-\, \frac{1}{\,4\,} \, C_3
-8 C_2
+ 6 R_{ai} \tau_a \sigma_i \,,
\end{equation}
where $C_3$ is the third order Casimir.
The relation directly follows by rewriting the
expression of $C_3$
(see Appendix. \ref{app;C3}).
As before,
the largest contribution comes from $C_3$
which appears to be $O(\Nc^3)$.
For the maximally symmetric spin-flavor wavefunction
of $n$ quarks, $C_3$ is
\begin{align}
C_3 \big( {\rm MS}(n) \big) = 24 \, n(n+4)(n-2) 
\,.
\end{align}
For the hedgehog and conventional wavefunctions,
we substitute $n=\Nc$,
and we get $\la C_3 \ra = 24\Nc(\Nc+4)(\Nc-2)$.

For the dichotomous wavefunction,
we have a sum of $C_3$'s for $(\Nc-1)$-quarks 
and an unpaired quark.
We get $24(\Nc-1)(\Nc+3)(\Nc-3) - 120
= 24( \Nc^3 - \Nc^2 - 9\Nc +4)$,
which differs from other two wavefunctions 
by $O(\Nc^2)$.

We already know the expectation values of $C_2$.  
The remaining $R \tau \sigma$ term yields
\begin{equation}
\la  R_{ai} \tau_a \sigma_i \ra 
=
\left\{
\begin{matrix}
 - \Nc^2  ~~~& ( {\rm hedgehog} )\\
~ 3( \Nc+2) ~~~& ~~~~({\rm conventional})\\
~ 9  ~~~& ~~~~({\rm dichotomous})
\end{matrix}
\right.
\end{equation}
The computation is given in Appendix.\ref{some}.
In particular, for the conventional and dichotomous wavefunctions,
we will see $\la R_{ai} \tau_a \sigma_i \ra = 9 | \la R_{33} \ra |$.

Assembling all these results,
the contribution from the $\omega-3\pi$ coupling 
is proportional to
\begin{equation}
- \epsilon_{abc} \epsilon_{ijk} \la  R_{ai} R_{bj} R_{ck} \ra 
= 6\Nc^3 + 6 \times
\left\{
\begin{matrix}
\, 7 \Nc^2 + 8\Nc ~~& ( {\rm hedgehog} )\\
\, 6\Nc^2 + 5\Nc - 6~~& ~~~~({\rm conventional})\\
\, 3\Nc^2 - \Nc + 3  ~~& ~~~~({\rm dichotomous})
\end{matrix}
\right.
\label{energyR3}
\end{equation}
after multiplying the overall sign for the potential energy.
The hedgehog state has the largest repulsive force.
Again we do not find the difference at
the leading order of $\Nc$.
Considering the overall factors of the 
Wess-Zumino vertex and the quark number density of $O(\Nc)$,
the energy difference appears to be
$ \sim \Nc g_{\omega qq}^2 /f_\pi^6 \times \Nc \times \Nc^2 
\sim \lqcd$.

Again we can explicitly see the size of $1/\Nc$ corrections.
The $1/\Nc$ corrections are substantial
as seen in $\la R_{ai}^2 \ra$.

\subsection{The hedgehog wavefunction
and mean field pions}

The hedgehog state has
special properties in processes with perturbative pions.
For linear $R$-operators,
it has large axial charge distributions
of $O(\Nc)$,
\begin{equation}
\la H | R_{33} | H \ra
= \la H | R_{+-} | H \ra
= \la H | R_{+-} | H \ra 
= - \Nc \,,
\end{equation}
or $\la R_{11} \ra = \la R_{22} \ra = \la R_{33} \ra = -\Nc$.
All the other expectation values of the $R$ operators vanish.

Let us first see the $R^2$-operator.
As we have seen,
the sum of the axial charge operator is
$\sim 3\Nc^2$.
It means that the hedgehog state well-saturates the 
contributions from the intermediate states:
\begin{equation}
\sum_{a,i} 
\la H | R_{ai}^2 |H \ra 
= 
\sum_{a,i} \sum_n
\big| \la H | R_{ai} |n \ra \big|^2 
\, \ge \,
\sum_{i=1}^{3} \big| \la H | R_{ii} | H \ra \big|^2 
= 3\Nc^2\,,
\end{equation}
where we used the hermiticity of $R_{ai}$.

The similar relation also holds for the
$R^3$-operator.
Its leading $\Nc$ contribution
is $- 6\Nc^3$,
which can be obtained by saturating the 
intermediate state with the hedgehog state
\begin{equation}
\epsilon_{ijk} \epsilon_{abc} 
\la H| R_{ai} R_{bj} R_{ck} | H \ra 
= 6 \la R_{11} \ra_H  \la  R_{22} \ra_H \la R_{33} \ra_H 
+ \cdots
= - 6\Nc^3 + \cdots,
\end{equation}
where $\cdots$ gives the remaining $O(\Nc^2)$ contributions.

As we have just seen,
for the hedgehog state,
the distribution of the axial charge operators
may be treated classically at large $\Nc$.
Correspondingly,
the mean field picture for pions is at work
for the hedgehog state.

In contrast,
usual baryon wavefunctions do not have
this factorization property,
and in fact, large contributions 
come from the transition amplitudes from
one baryon to other baryon states.
Thus quantum treatments are essential for these 
wavefunctions.
In particular, for the dichotomous wavefunction,
even if the expectation value of the linear operator $R$
(or the value of $g_A$)
is small,
the contributions from the off-diagonal matrix element
are very large.

\subsection{Color magnetic interaction}

Here we shall consider one gluon exchange
type processes
and see how it affects the energy splitting among
different wavefunctions.

The exchange of the spatial gluons (color-magnetic interaction)
provides the spin splitting potential,
($\lambda = \Nc g_s^2$, and $\vr_{qq'}$ is distance between
$q$-th and $q'$-th quarks)
\begin{equation}
\sum_{q \neq q'} V(\vr_{qq'})
 \sim \,
\frac{\lambda}{\Nc}
\sum_{q \neq q'} \frac{~ \sigma_i^{(q)} \sigma_i^{(q')} }{M_q M_{q'} } \,
\delta( \vr_{qq'} )\,, 
\end{equation}
which has been regarded as
a source of the $N-\Delta$ splitting in the 
constituent quark models\footnote{There is
some objection to this reasoning.
Liu and Dong performed a lattice simulation
with and without the $Z$ diagrams
which represent the meson exchanges.
After removing the $Z$ diagrams,
the $N-\Delta$ and $\pi-\rho$ are energetically
degenerate,
while the one gluon exchange should not be affected.
For details, see Ref.\cite{Liu:1998um}.
}.
Assuming all constituent quark masses are the same
and computing the spatial part
($\psi(\vr_{qq'})$: relative spatial wavefunction),
the matrix element can be written as
\begin{align}
\left\la \sum_{q \neq q'} V(\vr_{qq'}) \right\ra
& \sim \,
\frac{\, \lambda \, |\psi(\vec{0})|^2 \, }{\Nc M^2} \,
\left\la 
\left( \sum_{q} \sigma_i^{(q)} \right)^2
- \sum_{q} \left( \sigma_i^{(q)} \right)^2 
\right\ra 
\nonumber \\
&
= \frac{\, \lambda \, |\psi(\vec{0})|^2 \, }{\Nc M^2} \,
\left( \la \sigma_i^2 \ra - 3\Nc \right)
\,.
\end{align}
The second term is common for different wavefunctions,
so the difference comes from the first term.
For conventional (or dichotomous) 
wavefunction with fixed spin of $S=1/2$,
$\la \sigma_i^2 \ra = 3$, while
for the hedgehog state, 
$\la \sigma_i^2 \ra = 3\Nc$.
Since the interaction is repulsive,
the color magnetic interaction lifts up
the mass of the hedgehog state by $\sim \lqcd$,
compared to the conventional (or dichotomous) one.
This is a consequence that
the hedgehog state contains higher spin states.

\section{Non-perturbative analyses: 
Chiral Quark Soliton}
\label{Nonpert}

So far we have considered
the perturbative contributions,
provided that our wavefunctions
have the same constituent quark bases
with the same quark mass and quark-meson coupling.
Then we found that 
different wavefunctions give 
the same baryon self-energy at the leading $\Nc$.
However, once a large pion field is developed,
the quark bases can be modified,
and then the $O(\Nc^0)$ modification for each quark mass
may generate $O(\Nc)$ difference for a baryon mass.
We shall take into account such a possibility
within the framework of the chiral quark soliton model.

As we discussed in the previous section,
the static classical picture of the pion field
should work for the hedgehog state,
while it is difficult to apply this framework
for the conventional and dichotomous wavefunctions.
Therefore in this section,
we consider the hedgehog wavefunction only,
and discuss whether the generation of coherent pions 
is favored or not.

\subsection{A baryon with the pion mean field}

The chiral quark soliton model
is a model which smoothly interpolates
the constituent quark model
and topological soliton model.
We will work in the Euclidean space.
We start with the following Lagrangian
\begin{equation}
\calL = \bar{q} D(U_5 ) q \,,
~~~~
\big(\, D(U_5) = \left( \rmi \Slash{\partial} - M U_5 \right) \,,~
U_5 = \exp\left[ \rmi \gamma_5 \tau_a \pi_a \right] \, 
\big)\,,
\end{equation}
for which the regularized partition function is
\begin{equation}
Z 
= \int \calD U \calD q \calD \bar{q} \, \rme^{\, \rmi \int \rmd^4x \calL} 
\, \big|_{ {\rm reg.} }
= \int \calD U \exp\big[ 
\Nc \Tr \, \Ln \left( D( U_5) / D(U_5^0) \right)  \big]\,,
\end{equation}
where $U= \exp( \rmi \tau_a \pi_a )$.
Here we divide the partition function with general $U_5$
by that with the vacuum configuration, $U_5^0=1$.

For the vacuum,
the derivative expansion around $U_5^0=1$ generates
the non-linear $\sigma$ model plus
an infinite number of higher derivative terms,
that are powers of $\partial U_5/M$,
and the Wess-Zumino term 
\cite{Aitchison:1986aq}.
The generation of the coherent pions 
should cost energy.

The single quark bases for a given pion background 
is obtained
by solving the eigenvalue problem ($\tau=\rmi x^0$)
\begin{equation}
\gamma_0 D(U_5) \Phi_n
= \big( \! - \partial_\tau - h(U_5) \big) \Phi_n \,.
~~~~
\big(\, 
 h(U_5) = \gamma_0 (\, \rmi \gamma_k \partial_k + MU_5\,) \, 
\big)
\end{equation}
We will denote
the single particle energy 
for general $U_5$ and $U_5^0$ as
$\epsilon_{\alpha}$, $\epsilon_\alpha^0$, respectively.
These bases are used for the evaluation of the
quark determinant.

To compute the baryon spectra,
we must consider the baryonic correlator
at large time separation.
It is given by
\begin{align}
\hspace{-0.7cm}
\la J_B(\vec{0}, T) \bar{J}_B(\vec{0}, 0) \ra_{ U_5}
& \sim \,
\int \calD U
\exp\big[ \Nc \Tr \, \Ln \left( D(U_5)/D(U_5^0) \right) \big] \,
\prod_{i=1}^{\Nc} S_q(\vec{0}, T ;U_5)
\nonumber \\
& \sim \,
\int \calD U
\exp\big[ - T\,  
\big( E_{ {\rm sea} } [U_5] + E_{ {\rm val} } [U_5] \big) \big] \,,
\end{align}
where $E_{ {\rm sea} }$ comes from the regularized
Dirac determinant \cite{Diakonov:1987ty},
\begin{equation}
\Nc \Tr \, \Ln \left( D(U_5)/D(U_5^0) \right)
= - T \Nc \sum_{\epsilon_\alpha, \epsilon_\alpha^0 <0} 
( \epsilon_\alpha - \epsilon_\alpha^0 ) 
= -T E_{ {\rm sea} } (U_5)\,,
\end{equation}
and is $O(\Nc)$.
This part is considered to be responsible for the
polarization of the media that generates a pion cloud.
Actually even after subtracting the trivial vacuum contribution,
the sea contribution still diverges
and has to be regularized by the UV 
cutoff\footnote{In a more realistic treatment,
there should be a form factor effect 
for the quark-pion coupling,
which is likely to remove the sensitivity to the 
UV cutoff.}.
The coherent pion field affects quarks in the Dira sea
with large negative energies.

On the other hand, 
the $E_{ {\rm val} }$ is the valence quark contributions
originating from the interpolating fields
(the product of $\Nc$-propagators),
\begin{equation}
E_{ {\rm val} } 
= \sum_{\alpha=1}^{\Nc} 
\epsilon^\alpha_{ {\rm val} } ( U_5) \,, ~~~ 
( \epsilon_{ {\rm val} } (U_5): {\rm the~level~above~the~Dirac~sea})
\end{equation}
which is also $O(\Nc)$.
Here the $\epsilon_{ {\rm val} }$ must appear
above the Dirac sea because of the Pauli-blocking.

While the formal expression of the baryon mass
is rather simple, its computation is quite tough.
The derivative expansion around the vacuum configuration
is hard to control,
because the pion configuration varies at the scale of $\lqcd^{-1}$.
It means that all the higher powers of $\partial U_5/M$ equally contribute,
so that it is necessary to
completely diagonalize the quark determinant as well as
the valence quark propagator.
Such treatments have been carried out
numerically in seminal 
works \cite{Diakonov:1987ty,Reinhardt:1988fz,Alkofer:1994ph}.
We will borrow their results for our discussions.

\subsection{The optimized configuration
within the stationary phase approximation}

Note that the valence quark contributions are $O(\Nc)$,
thus can affect the optimal configuration of $U_5$ at large $\Nc$.
The $U_5$ may deviate from $1$ at the classical level.
We determine the optimal
configuration $U_5$ from the following equation,
\begin{equation}
\delta_{U_5} 
\big( E_{ {\rm sea} } [U_5] + E_{ {\rm val} } [U_5] \big) 
\big|_{U_5 = U_5^c} =0 \,.
\label{minima}
\end{equation}
For two flavor case, we may parameterize $U_5$ as
\begin{equation}
U_5 
= \exp\big[ \rmi \gamma_5 \Theta(\vx) n_a (\vx) \tau_a \big]
= \cos\Theta(\vx) 
+ \rmi \gamma_5 n_a (\vx) \tau_a \sin\Theta(\vx) \,,
\end{equation}
where $n_a^2=1$.
It is related to the linear realization as
\begin{equation}
\sigma(\vx) = M \cos\Theta(\vx) \,,~~~
\pi_a(\vx) =  n_a(\vx) M \sin\Theta(\vx) \,.
\end{equation}
The $\Theta(\vx)$ is responsible for the
fraction of the chiral density and pion density,
while $n_a(\vx)$ characterizes
orientations of pions in the isospin space.

In Eq.(\ref{minima}),
if we take the variation with respect to $\Theta(\vx)$,
it yields
\begin{equation}
\sin\Theta(\vx) \, 
\big\la \bar{q} q (\vx) \big\ra 
- 
n_a(\vx) \cos\Theta(\vx) \, 
\big\la \bar{q} \rmi \gamma_5 \tau_a q (\vx) \big\ra = 0\,. 
\label{self}
\end{equation}
Since the expectation value of the operators
depend on the quark bases,
this is the self-consistent equation.

\subsection{The behaviors of $E_{ {\rm field}}$ and $E_{ {\rm val} }$ }

Consider first the trivial configuration 
$U_5=1 \, (\Theta=0)$
for which $E_{ {\rm sea} }=0$ by definition.
Then the baryon mass is determined
by the lowest quark energy orbit without a pion cloud,
and $M_B=\Nc M$ at this level of the treatment.
Thus if a nontrivial configuration of $U_5$ would exist,
at least it must give smaller baryon mass than $\Nc M$.

Now suppose that the hedgehog quark 
wavefunction would give the ground state.
Through the Dirac equation for the 
quark hedgehog wavefunctions \cite{Birse:1984js},
the form of $n_a(\vx)$ is fixed to
$n_a(\vx) = r_a/r$,
and $\Theta(\vx) = \Theta(r)$
where $\vr =\vx -\vx_0$ 
($\vx_0$: the center of the baryon).

For this configuration,
the boundary condition at $r \rightarrow0$
is imposed to avoid
the singularity related to the angular variable $n_a$
or $\pi_a$.
It requires $\Theta(r) \rightarrow  n\pi$ ($n$: integer)
or $\pi_a (r) \rightarrow 0$ as $r\rightarrow 0$.
The integer $n$ appears to correspond to the
topological number of the meson profile.
On the other hand
$\Theta(r) \rightarrow 0$ as $r\rightarrow \infty$
to recover the vacuum configuration
at asymptotic distance.

We have to solve the self-consistent equation (\ref{self})
numerically.
Instead,
we use the profile function
\begin{equation}
\Theta(r;R) = - \pi \exp( - r/R) \,,
\label{profile}
\end{equation}
to see the general tendency of a single
particle spectrum and the Dirac sea contributions
in the presence of the hedgehog background.
Actually, the self-consistent solution of Eq. (\ref{self})
is known to give the similar form \cite{Reinhardt:1988fz}.
Here $R$ is treated as a variational parameter.

Clearly $E_{ {\rm sea} }$ is a monotonously 
increasing function of $M$ or $R$.
Meson profiles with larger $R$ and $M$ 
just increase the energy cost
to make a hole in the vacuum condensate.
On the other hand, the single particle energy $E_{ {\rm val } }$
is a monotonously decreasing function of $M$ or $R$.
The couplings between quark and mesons 
are attractive,
so larger $M$ gives larger energy reduction.
For larger $R$, the quark can be more deeply
bound to the mesons, because of the
smaller kinetic energy in the potential with broader range.
At $MR \gg 1$,
the valence energy level even enters
the negative continuum,
approaching to $-M$.

\begin{figure}[tb]
\vspace{0.0cm}
\begin{center}
  \includegraphics[scale=.25]{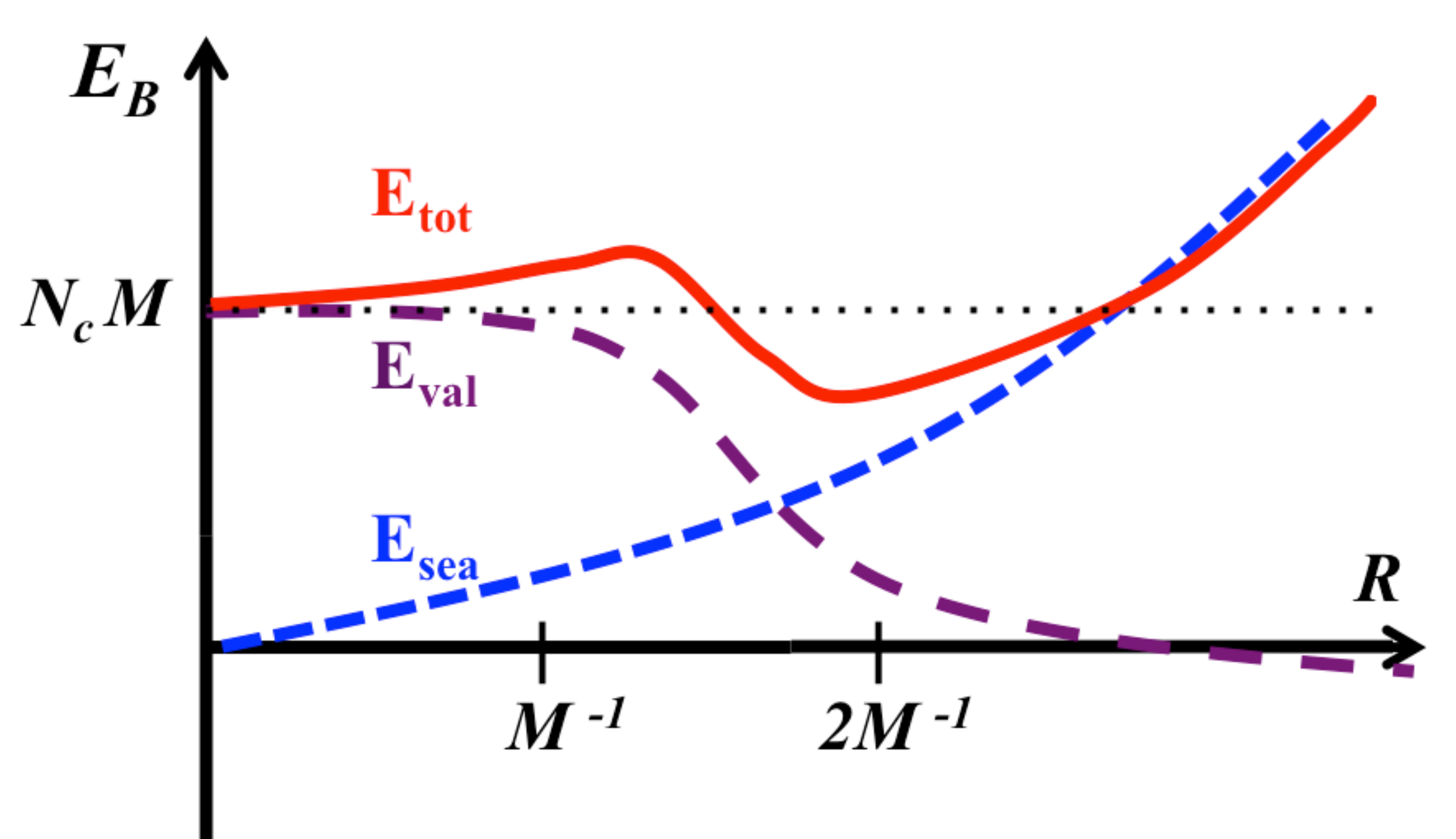} 
\end{center}
\vspace{0.0cm}
\caption{A schematic plot of 
$E_{ {\rm val}}$, $E_{ {\rm sea}}$, and their sum
$E_{ {\rm tot} }$ as a function of a meson cloud size $R$
in the profile (\ref{profile}).
The decreasing behavior of $E_{ {\rm val} }$
is due to a larger binding energy for a broader potential well.
The increasing behavior of $E_{ {\rm sea} }$
comes from the Dirac sea polarization.
In $E_{ {\rm tot} }$,
the minimum is typically found at
some value of $R$ between $M^{-1}$ and $2M^{-1}$,
but the details about the height and locations
of the minimum depend on the UV 
cutoff for the model.
Here we took the UV cutoff to make the sea contribution
small enough.
With too large UV cutoff,
the curve for $E_{ {\rm sea} }$ is lifted up, making
the minimum a local one.
}
\vspace{0.2cm}
\label{fig:level}
\end{figure}

These tendencies are summarized in Fig.\ref{fig:level}
for our trial meson profile (\ref{profile}).
The minimum (or local minimum)
typically appear around $R \simeq (1-2) M^{-1}$
after cancelling out very large binding energy
and sea quark contributions.
The corresponding baryon
is a mixture of valence quarks and coherent pions.

Actually whether the minimum appears
below or above depends on subtle details in calculations,
such as the UV cutoff.
The larger UV cutoff increases $E_{ {\rm sea} }$
because in this model
coherent pions even affect the energy level
very deep in the Dirac sea,
costing energy \cite{Reinhardt:1988fz}.
But in what follows, the existing literatures
suggest that
even if we find a nontrivial minimum with a coherent pion cloud,
the energy reduction from $\Nc M$
is $20-30$\% at most.

\subsection{Spatial size of coherent pion and
confining effects}

%
\begin{figure}[tb]
\vspace{0.0cm}
\begin{center}
  \includegraphics[scale=.25]{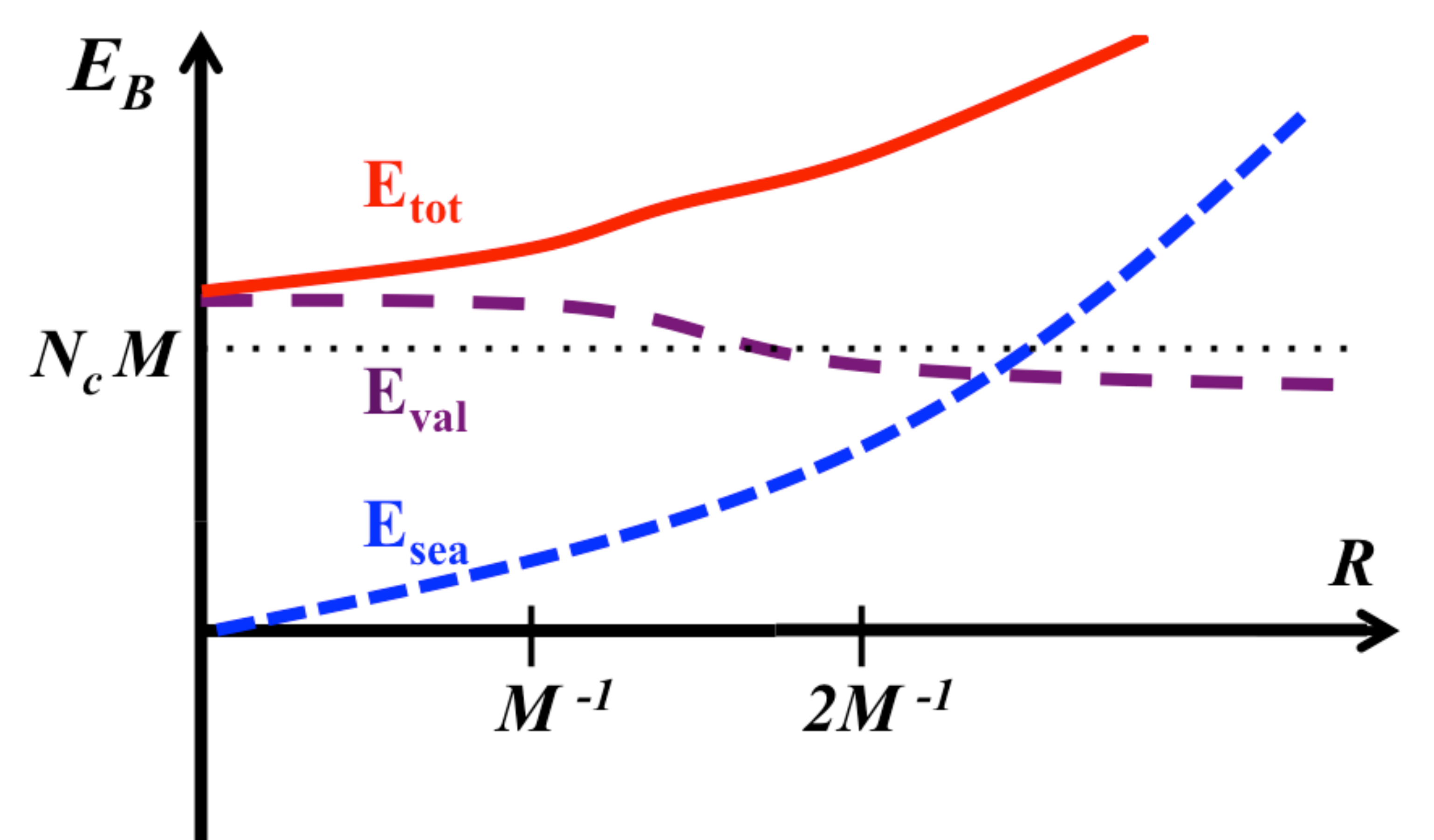} 
\end{center}
\vspace{0.0cm}
\caption{An expectation when the confining effects
are included for the previous plot in Fig.\ref{fig:level}.
(The confining potential lifts up the quark kinetic energy
so even at $R\sim 0$,
so the energy is higher than $\Nc M$.)
The main modification appears in the behavior of
$E_{ {\rm val}}$.
At distance scale $r \sim M^{-1}$,
the linear potential closes a potential well made of a pion cloud,
so the valence level does not acquire energy reduction much.
As a consequence,
the local minimum in $E_{ {\rm tot} }$ would disappear.
}
\vspace{0.2cm}
\label{fig:levelconf}
\end{figure}

The reduction of the baryon energy
only comes from the quark binding to a pion cloud.
The spatial size of the pion cloud must be
so large, $R \ge M^{-1}$, that
bound quark wavefunctions can
spread out to avoid the kinetic energy cost.

However, the quark wavefunction cannot be
widely spread out once confining effects are turned on.
Our expectation is summarized in Fig.{\ref{fig:levelconf}.
The confining force is activated when 
a quark is at distance of $r \sim \lqcd^{-1}$
from the center of the baryon,
giving the energetic cost of $\sim \sigma r \sim \lqcd$.
When we add contributions from
the pion cloud potential and
the linear potential,
the potential well in the former with depth $\sim M$ 
is always closed at distance $r \sim M^{-1}$
by the linear potential,
no matter how $R$ is large.
Then the deeply bound valence quarks
may not be allowed even at very large $R$.

As a consequence, it is possible that
the nontrivial minimum, 
that was found within the chiral quark soliton model,
disappears by including confining effects.
As we have mentioned,
the energy reduction by the quark binding to pions
is not so large, so it is not surprising
that confining effects lift the energy of 
the baryon with coherent pions 
above those without coherent pions.
If this is indeed the case,
the true energy minimum should be found at
the small size of the coherent pion, $R \ll M^{-1}$.
Then a number of pions becomes small,
invalidating the mean field or coherent pion picture.
We have to go back to the quantum calculations 
to treat quantum pions rather than coherent pions.

\section{Summary}
\label{Discussion}
In this paper, we have discussed
the energy differences among
the hedgehog, conventional, and
the dichotomous nucleon wavefunctions.
Mainly we focus on the self-energy difference
originating from the pion exchange
which is related to the matrix element of the 
axial charge operators.

After examining both the perturbative and non-perturbative
treatments of pions,
we realize that it is very hard to derive 
decisive conclusions on the lowest energy state.
The perturbative treatments
give the same leading $\Nc$ behaviors for 
our three wavefunctions.
In the non-perturbative treatments within 
the chiral quark soliton models,
the existence of the state with coherent pions
seems to be a subtle issue
and depends upon details such as the UV cutoff
and the confining force which 
are difficult to include in a solid way.

On the other hand, there seems to be
a general tendency that
both one gluon exchange and confining interaction
generate extra energy costs for the
hedgehog wavefunction,
which is a representative state with
coherent pions.
Both effects are not manifest in 
typical chiral effective models. 
The combination of these effects with
chiral models deserves further studies.

Clearly our analyses are incomplete.
We dropped off exchanges of higher towers of mesons
that couple to the axial charges.
We expect that such resonances just change
an overall size of the interactions,
but do not alter the ordering of three wavefunctions.
We have to check more carefully 
whether this expectation is correct or not, though.

Another oversimplification is the non-relativistic treatments.
In particular,
it is often said that at $\Nc=3$,
the non-relativistic value of the axial charge is $5/3$ 
for the conventional nucleon,
but it can be reduced to the experimental value $\simeq 1.26$
by including the down component of the Dirac spinor.

However, presumably this is not the end of the whole story.
The nucleon $g_A$ was
measured on the lattice for
several pion masses 
\cite{Edwards:2005ym,Yamazaki:2008py,Capitani:2012gj}.
The value is typically lower than the experimental value
by about $10$\%.
What is nontrivial is
that even at a large pion mass $\simeq 700$ MeV,
$g_A$ remains small, $\simeq 1.1-1.2$.
The non-relativistic treatments 
should work better for  larger $m_\pi$
and higher constituent quark mass,
reducing corrections from pion dynamics and
relativistic effects.
Thus we need explanations
other than the relativistic corrections.

As for the lattice studies,
it should be also mentioned that
a lattice study of the baryon spectra is recently carried out
for $\Nc=3,5,7$,
in which the lowest pion mass is $(m_\pi/m_\rho)^2 \simeq 0.2$
(exp.: $\simeq 0.03$)
\cite{DeGrand:2012hd}.
The results show the rigid rotor spectra
for the excited baryons,
that are consistent with predictions
from the color magnetic interaction,
the perturbative pion exchange,
and the collective excitations of the chiral soliton.
The energy splitting is $\sim 1/\Nc$.
Again we cannot judge which predictions are preferred,
but we hope that eventually detailed studies of the pion mass 
dependence will distinguish these three contributions.

This lattice study also brought
a serious constraint on the dichotomous wavefunction.
Once we assumed the ground state with $I=S=1/2$ to be
the small $g_A$ nucleon,
it is natural to assume that
the ground state with $I=S=3/2$ is
the $\Delta$ baryon with small $g_A$.
To construct such a $\Delta$ state,
we need to bring two valence quarks 
into extra spatial orbits, costing energy of $\sim \lqcd$
instead of $\sim \lqcd/\Nc$ \cite{Hidaka:2010ph}.
This contradicts with the lattice data
for relatively large pion mass,
while the dichotomous model is originally proposed
to avoid a large pion cloud whose effects
become substantial for small pion mass.
The lattice studies for smaller pion mass
may provide clear judgment for 
the validity of the dichotomous picture.

Finally we have to admit that
the $1/\Nc$ corrections are substantial, at least
in the perturbative estimates of the self-energy contributions
(see the expressions, (\ref{R2Nc}) and (\ref{energyR3})).
This fact may suggest us to revise
our original proposal (see introduction),
by taking a more appropriate
leading order of the expansion.

Maybe a part of the problem
comes from the fact that
we put a large number of quarks into a small flavor space
(two-flavor).
It is interesting to check
what happens if we take the Veneziano limit, $\Nc,\Nf \rightarrow \infty$,
$\Nf/\Nc=$ fixed.
For instance, the second Casimir 
for the maximally symmetric representation is given by
\begin{equation}
C_2 = (2\Nf-1) \times \Nc (\Nc+ 2\Nf) \,.
\end{equation}
It is clear that the $1/\Nc$ correction appears as
the $\Nf/\Nc$ correction, and is substantial.
Other quantities and their qualitative impacts
on the nuclear physics
will be discussed elsewhere.

\section*{Acknowledgments}
This work is an extension of the previous work
with Y.~Hidaka, L.~McLerran and R.~D.~Pisarski.
T.K. acknowledges them for many insightful discussions.
Special thanks go to L. McLerran for encouragements
and carefully reading the manuscript.
T.K. is supported by
Humboldt foundation through
its Sofja Kovalevskaja program.

\appendix

\section{More on the $SU(4)$ algebra}

For later calculations, 
we wish to write commutation relations in terms of the
Cartan bases \cite{Lie}.
We choose the following Cartan generators
$(H_1, H_2, H_3) =(\tau_3\,,  \sigma_3\,,  R_{33})$.
To find other bases, we must take
appropriate linear combinations of 
$\tau$, $\sigma$, and $R$ to find generators $E_{\pm \alpha}$
such that
\begin{equation}
[\, H_i \,, E_{\pm \alpha} \,] = \pm \, \alpha_i E_{\pm \alpha} \,.
\end{equation}
Let us find such linear combinations.
First we define
\begin{align}
&\hspace{-0.3cm}
\tau^{(q)}_{\pm} = \frac{\tau^{(q)}_1 \pm \rmi \tau^{(q)}_2}{2}\,,
~~~~
\sigma^{(q)}_{\pm} = \frac{\sigma^{(q)}_1 \pm \rmi
  \sigma^{(q)}_2}{2}\,,
\nonumber \\
&\hspace{-0.3cm}
R^{(q)}_{3 \pm} = \tau^{(q)}_{3} \sigma^{(q)}_{\pm} \,,
~~
R^{(q)}_{\pm 3} = \tau^{(q)}_{\pm} \sigma^{(q)}_{3} \,,
~~
R^{(q)}_{\pm \pm} = \tau^{(q)}_{\pm} \sigma^{(q)}_{\pm} \,,
~~
R^{(q)}_{\pm \mp} = \tau^{(q)}_{\pm} \sigma^{(q)}_{\mp} \,.
\end{align}
Not all of these generators may be directly used as $E_{\pm \alpha}$.
In particular, $\sigma_\pm$ and $\tau_\pm$ operators
can not be used as raising and lowering operators
for the irreducible representations of $SU(4)$,
because states with definite $R_{33}$ eigenvalues
are mixtures of states with different spins and isospins.
It turns out that our bases are
\begin{align}
& 
E_{\pm \alpha^{(1)} } = R_{\pm 3} + \tau_\pm\,,
~~~
E_{\pm \alpha^{(2)} } = R_{\pm 3} - \tau_\pm \,,
~~~
E_{\pm \alpha^{(3)} } = R_{3 \pm} + \sigma_\pm\,,
\nonumber \\
& 
E_{\pm \alpha^{(4)} } = R_{3 \pm} - \sigma_\pm \,,
~~~
E_{\pm \alpha^{(5)} } = 2R_{\pm \pm} \,,
~~~
E_{\pm \alpha^{(6)} } = 2R_{\pm \mp} \,.
\label{generators}
\end{align}
Here the root vectors are
\begin{align}
&
\vec{\alpha}^{(1)} = ( 2, 0, 2) \,,
~~~
\vec{\alpha}^{(2)} = ( 2, 0, -2) \,,
~~~
\vec{\alpha}^{(3)} = ( 0, 2, 2) \,,
\nonumber \\
&
\vec{\alpha}^{(4)} = ( 0, 2, -2) \,,
~~~
\vec{\alpha}^{(5)} = ( 2, 2, 0) \,,
~~~
\vec{\alpha}^{(6)} = ( 2, -2, 0) \,,
\end{align}
which give the relation
\begin{equation}
[E_{\alpha^{(j)} } , E_{-\alpha^{(j)} } ] = \alpha_i^{ (j) } H_i \,.
\end{equation}
%

\section{The evaluation of Casimir operators}

Here we briefly summarize the construction
of the Casimir operators
for the Lie algebra.
To derive the general form,
we follow the discussions of Okubo \cite{Okubo}.
Then we use it to explicitly derive
the formula for the maximally symmetric (MS) 
representations.

\subsection{The general form}
We consider generators of the Lie algebra
\begin{equation}
[X_a, X_b] = f_{ab}^{~~c} X_c\,.
\end{equation}
The general form of the Casimir operator is given by
($\tilde{C}_p$ can differ from $C_p$ by an overall factor
which will be fixed later)
\begin{equation}
\tilde{C}_p = g^{a_1 a_2 \cdots a_p} X_{a_1} X_{a_2} \cdots X_{a_p} \,,
\end{equation}
where $X_{a_1}, \cdots, X_{a_p}$, are generators of the group,
and
\begin{equation}
g_{ \{\lambda_0\} }^{a_1 a_2 \cdots a_p} 
= \tr_{\lambda_0} 
\left[ X^{a_1}_{ \{\lambda_0 \} } X^{a_2}_{ \{\lambda_0\} } 
\cdots X^{a_p}_{ \{\lambda_0 \} } \right] \,.
\end{equation}
Here $X_{ \{\lambda_0\} }$ means a generator in a
representation $\{\lambda_0\}$ which we shall
call a reference representation.
For $\{\lambda_0\}$,
we use the fundamental representation of $SU(2\Nf)$
with $\tr_{ {\rm fund.} } [1] =4$,
and we raise and lower the indices using
$g_{ab} = 2\Nf \delta_{ab}$ and 
$g^{ab} = (g^{-1})_{ab} = \delta_{ab}/2\Nf$.
We sometimes drop off the subscript $\{\lambda_0 \}$
in $g^{a_1 \cdots a_p}$ to simplify equations.

We wish to evaluate the value of the Casimir 
for any representations.
Actually, once we compute the value of the quadratic
Casimir for given representations,
we can express the higher order Casimir using $C_2$'s.
Below we briefly explain
a main idea of the derivation.

Let us consider a product space
$\{\lambda\} \otimes \{ \lambda_0 \}$,
and its decomposition into irreducible representations,
\begin{equation}
\{ \lambda \} \otimes \{\lambda_0 \}
= \sum_{j=1}^N \oplus \{ \lambda_j \} \,.
\end{equation}
In the RHS, there can be degeneracies of representations,
for instance, it is possible that 
$\{ \lambda_k \} = \{\lambda_l \}$ for $k\neq l$.
We wish to compute the Casimir for 
the representation $\{\lambda \}$,
using the fundamental representation for 
$\{ \lambda_0 \}$.

Consider a generator in the product space
and its decomposition to irreducible representations,
\begin{equation}
X_a^{ \{ \lambda \otimes  \lambda_0 \}  } 
= X_a^{ \{ \lambda \} } \otimes E^{ \{ \lambda_0 \}} 
+ E^{ \{ \lambda \}} \otimes X_a^{ \{ \lambda_0 \}}
= 
\sum_{j=1}^N X_a^{ \{ \lambda_j \} } P_j \,,
\end{equation}
where $P_j$ is a projection operator
picking out the $j$-th representation space.
Clearly 
$\sum_j P_j = E^{ \{ \lambda \}} \otimes E^{ \{ \lambda_0 \}}$.

Now let us define the following operator
\begin{equation}
Q_{ \lambda; \lambda_0} 
\equiv X_a^{ \{ \lambda \} } \otimes  X^a_{ \{ \lambda_0 \}} 
= g^{ab} X_a^{ \{ \lambda \} } \otimes  X_b^{ \{ \lambda_0 \}} 
\,,
\end{equation}
which commutes with all the generators
in $\{ \lambda \otimes  \lambda_0 \}$,
\begin{equation}
\left[ Q_{\lambda; \lambda_0} \,,\,  X_a^{ \{ \lambda \otimes  \lambda_0 \}  } \right] = 0 \,.
\end{equation}
Using this property,
we can further show that the $p$-th order product
\begin{equation}
( Q_{\lambda;\lambda_0} )^p 
= X_{a_1}^{ \{ \lambda \} }  X_{a_2}^{ \{ \lambda \}} \cdots X_{a_p}^{ \{ \lambda \}}  
\otimes  
X^{a_1}_{ \{ \lambda_0 \} }  X^{a_2}_{ \{ \lambda_0 \}} \cdots X^{a_p}_{ \{ \lambda_0 \}}  
\,,
\end{equation}
commutes with all the generators,
and is the Casimir operator in the product space
$\{ \lambda \otimes  \lambda_0 \}$.
Now taking the partial trace for $\{\lambda_0 \}$,
we obtain $g^{a_1 \cdots a_p}$.
Then the operator $\tr_{\lambda_0} [ Q^p ]$ gives
the Casimir $\tilde{C}_p (\lambda)$ 
for the representation $\{\lambda\}$.

The relation between the $\tilde{C}_p(\lambda)$ 
and $\tilde{C}_2$'s
follows from the property of the operator 
$Q_{\lambda; \lambda_0}$.
We can rewrite $Q_{\lambda; \lambda_0}$ as
a sum of $\tilde{C}_2$ for the representations 
$\{ \lambda_j \}$, $\{\lambda\}$, and $\{\lambda_0\}$,
\begin{align}
 Q_{\lambda; \lambda_0} 
&= g^{ab} \, X_a^{ \{ \lambda \} } \otimes  X_b^{ \{ \lambda_0 \}} 
\nonumber \\
&= \frac{g^{ab}}{2} \left[
\left( X_a X_b \right)^{ \{\lambda \otimes \lambda_0 \} }
- \left( X_a X_b \right)^{ \{\lambda \} } \otimes E^{ \{ \lambda_0 \}
}
- E^{ \{ \lambda \} } \otimes \left( X_a X_b \right)^{ \{\lambda_0 \} } 
\right] 
\nonumber \\
&= \frac{1}{2} \, \sum_{j=1}^N
\big[
\tilde{C}_2(\lambda_j) - \tilde{C}_2(\lambda) - \tilde{C}_2(\lambda_0)
\big] P_j 
\,.
\end{align}
Below we will write
$\xi_{ \lambda_j; \lambda, \lambda_0} 
= 
\big[ \tilde{C}_2(\lambda_j) 
- \tilde{C}_2(\lambda) - \tilde{C}_2(\lambda_0) \big]/2$.
Then the $p$-th order product is
\begin{equation}
( Q_{\lambda;\lambda_0} )^p 
= X_{a_1}^{ \{ \lambda \} } \cdots X_{a_p}^{ \{ \lambda \}}  
\otimes  
X^{a_1}_{ \{ \lambda_0 \} }  \cdots X^{a_p}_{ \{ \lambda_0 \}}  
=
\sum_{j=1}^N \left( \xi_{\lambda_j; \lambda, \lambda_0} \right)^p P_j \,.
\end{equation}
Now a full trace of the operator $(Q_{\lambda; \lambda_0} )^p$
for $\{ \lambda \}$ and $\{ \lambda_0 \}$
gives the $p$-th order Casimir operator for $\{ \lambda \}$
times the dimension of $\{ \lambda \}$
(or the Casimir operator for $\{ \lambda_0 \}$
times the dimension of $\{ \lambda_0 \}$).
The full trace in RHS gives a sum of the dimension for
each representation $\{ \lambda_j \}$.
Therefore we finally find that
\begin{equation}
\tilde{C}_p(\lambda) 
= \sum_{j=1}^N \left( \xi_{\lambda_j; \lambda, \lambda_0} \right)^p
\, \frac{ d(\lambda_j) }{ d(\lambda) } \,,
\end{equation}
where $d(\lambda)$ is the dimension of the 
representation $\{ \lambda \}$.
To compute $\tilde{C}_p(\lambda)$,
we need to know $\tilde{C}_2$'s of all the irreducible representations
out of the product of 
$\{ \lambda \} \otimes \{ \lambda_0 \}$.

\subsection{The Casimirs for the maximally 
symmetric representations}

In the following we will consider the Casimir for the
MS representation for $n$-quarks.
We need to compute
the product of the MS($n$) representation and
the fundamental representation
which gives only two irreducible representations.
In the Young tableaux,
\begin{equation}
\YoungTab[0][]{ {,,,,} } \cdots \YoungTab[0][ ]{ {,, } } 
~~~\otimes ~~\YoungTab[0][]{ { } }
~~ = ~~
 \YoungTab[0][]{ {,,,,} } \cdots \YoungTab[0][ ]{ {,,, } } 
~~~ \oplus ~~
\YoungTab[-1][]{ {,,,,} { } } \cdots \YoungTab[0][ ]{ {,, } } 
\end{equation}
In the RHS,
the first one is the MS wavefunction of $(n+1)$-quarks,
and the other is the mixed symmetric representation,
which we will denote ``mS''.

The dimension of the MS representation
for $n$-quarks is given by
\begin{equation}
d \big( {\rm MS} (n) \big) 
= \frac{ \, ( n + 2\Nf -1)! \, } { n! \, (2\Nf-1) ! } \,.
\end{equation}
By comparing both sides of the Young tableaux,
we can find the dimension of the mS wavefunction with 
$(n+1)$-quarks as
\begin{align}
d \big( {\rm mS} (n+1) \big) 
&= d \big( {\rm MS} (n) \big) \times 2\Nf 
- d \big( {\rm MS} (n+1) \big)  
\nonumber \\
&= \frac{ \, n \, ( n + 2\Nf -1)! \, } { (n+1) ! \, (2\Nf-2) ! } 
\,.
\end{align}
Since
\begin{equation}
\frac{ d \big( {\rm MS} (n+1) \big) }{ d \big( {\rm MS}(n) \big) }
= \frac{\, n + 2\Nf \, }{ n+1 } \,,~~~~~
\frac{ d \big( {\rm mS} (n+1) \big) }{ d \big( {\rm MS} (n) \big) }
= \frac{\, n \, (2\Nf -1) \, }{\, n+1 \,} \,,
\end{equation}
we find
\begin{equation}
\tilde{C}_p \big( {\rm MS} (n) \big) 
= \left( \xi_{ {\rm MS}(n+1) } \right)^p
\frac{\, n + 2\Nf \,}{ n+1 } 
+
\left( \xi_{ {\rm mS}(n+1) } \right)^p
\frac{\, n \, (2\Nf -1)  \,}{\, n+1 \,} 
\,.
\end{equation}
The remaining task is the evaluation of the
quadratic Casimir for the MS and mS representations.

\subsection{The second order Casimir
for MS and mS representations}
\label{Casimir2}

The quadratic Casimir can be
written in terms of the Cartan bases as follows:
\begin{equation}
C_2 = \sum_{i=1}^{3} H_i^2
+ \sum_{j=1}^{6} \{ E_{\alpha^{(j)} } \,, E_{ -\alpha^{(j)} } \} \,.
\end{equation}
This is related to $\tilde{C}_2 = C_2/4$.
($\tilde{C}_2=X_aX^a = g^{ab} X_a X_b = \delta^{ab} X_a X_b/4$.)
We will evaluate it for the MS and mS representations
with $n$-quarks.

For the MS representation,
we need to consider only the state with the highest weight,
$( | \uup \ra )^n = | n, n, n \ra$.
In the second sum,
$E_{\pm \alpha^{(j)} } |n,n,n\ra=0$
for $j=2,4,6$,
so 
only terms with $j=1,3,5$ can contribute for this state.
Since $\vec{\alpha}^{(1)} = (2,0,2)$, 
$\vec{\alpha}^{(3)} = (0,2,2)$, 
and $\vec{\alpha}^{(5)} = (2,2,0)$,
we have
\begin{equation}
C_2 |n,n,n\ra 
= 
\left( \sum_{i=1}^{3} H_i^2
+ 4( \tau_3 + \sigma_3 + R_{33} )  
\right)
|n,n,n \ra
= 3n (n+4) |n,n,n\ra
\,,
\end{equation}
where we rewrite 
$\{ E_\alpha , E_{-\alpha} \} = [E_\alpha, E_{-\alpha} ] +
2E_{-\alpha} E_{\alpha}$,
and use the relation
$[E_{ \alpha^{ (j) } }, E_{-\alpha^{ (j) }} ] = \alpha_i^{ (j) } H_i$.

Next we consider the mS wavefunction
with $(n+1)$-quarks.
First we prepare the highest weight wavefunction
of the mS representation
which is orthogonal to the MS wavefunction
with the same $(\tau_3, \sigma_3, R_{33})$ values.
Consider the MS wavefunction
with the weight of the highest value minus one root,
\begin{equation}
| n-1, n-1, n+1 \ra_{{\rm MS}}
\, \propto \, 
\calS \, [\, |\uup\ra^n | \ddown \ra ] \,,
\end{equation}
which can be rewritten as
\begin{align}
&\calS \, \big[\, |\uup\ra^n | \ddown \ra \big]
= 
\calS \, \big[\, |\uup\ra^n \big] \otimes  | \ddown \ra 
+ n\, 
 \calS \, \big[\, |\uup\ra^{n-1} | \ddown \ra \big] 
\otimes | \uup \ra 
\nonumber \\
& = n! \, \left( \, 
| n, n, n \ra_{{\rm MS}} \otimes  | \ddown \ra 
+ \sqrt{n} \,
| n-2, n-2, n \ra_{{\rm MS}} \otimes  | \uup \ra 
\right) \,.
\end{align}
Then the mS wavefunction can be constructed
as follows
\begin{align}
&
| n-1, n-1, n+1 \ra_{{\rm mS}}
\nonumber \\
& 
= \frac{1}{ \sqrt{ n+1\, } \,}
\big(
\sqrt{n} \, | n, n, n \ra_{{\rm MS}} \otimes  | \ddown \ra 
- 
| n-2, n-2, n \ra_{{\rm MS}} \otimes  | \uup \ra 
\big) \,.
\end{align}
To evaluate the Casimir value,
we decompose the generators as follows:
\begin{equation}
X_a^2 = \left( \sum_{q=1}^n X_a^{(q)} + X_a^{(q')} \right)^2
= \left( \sum_{q=1}^n X_a^{(q)} \right)^2 
+ 2 \sum_{q=1}^n X_a^{(q)} X_a^{(q')} 
+ ( X_a^{(q')} )^2  \,,
\label{tosee}
\end{equation}
where the $X_a^{(q')}$ acts on the $(n+1)$-th quark.
The square operators give the Casimir values
for the MS representation for $n$-quarks
and a single quark.
Nontrivial contributions
come from the product of $n$-quark operators and
a single quark operator, which 
can be rewritten as
\begin{equation}
\sum_{q=1}^n X_a^{(q)} X_a^{ (q')} 
= \sum_{q=1}^n \left(
H_i^{(q)} H_i^{ (q')} 
+ \sum_{j=1}^6 \left(
E_{\alpha^{(j)} }^{(q)} E_{-\alpha^{(j)}}^{(q')} 
+
E_{-\alpha^{(j)} }^{(q)} E_{\alpha^{(j)}}^{(q')} 
\right)
\right) \,.
\end{equation}
When these product part act on the 
state $| n, n, n \ra_{{\rm MS}} \otimes  | \ddown \ra$,
only $H_i$ and $E_{\alpha^{(5)}}$ parts
give the nonzero contributions.
We find 
\begin{align}
& \sum_{q=1}^n X_a^{(q)} X_a^{ (q')} 
| n, n, n \ra_{{\rm MS}} \otimes  |\ddown \ra
\nonumber \\
&= - n \, | n, n, n \ra_{{\rm MS}} \otimes  |\ddown \ra
+ 4 \sqrt{n} \,
 | n-2, n-2, n \ra_{{\rm MS}} \otimes  |\uup \ra \,,
\end{align}
where we use the relation 
$\sum_q E^{(q)}_{-\alpha^{(5)}} |n,n,n\ra 
= 2 \sqrt{n} \, |n-2,n-2,n\ra$
and $E^{(q')}_{\alpha^{(5)}} |\ddown \ra = 2 \, | \uup \ra$.
Similar calculations are applied to 
$| n-2, n-2, n \ra_{{\rm MS}} \otimes  | \uup \ra$,
for which only $H_i$ and $E_{\alpha^{(5)}}$ survive.
After summing these two contributions,
one finds 
\begin{equation}
\sum_{q=1}^n  X_a^{(q)} X_a^{ (q')} 
| n-1, n-1, n+1 \ra_{{\rm mS}} 
= - (n+4)\, | n-1, n-1, n+1 \ra_{{\rm mS}} \,,
\end{equation}
so that $C_2 \big( {\rm mS}(n+1) \big)$ is given by
\begin{align}
&
\left[ \left( \sum_{q=1}^n X_a^{(q)} \right)^2 
+ 2 \sum_{q=1}^n X_a^{(q)} X_a^{(q')} 
+ ( X_a^{(q')} )^2  \right] \, |n-1,n-1,n+1\ra_{{\rm mS}}
\nonumber \\
&= \big[ \, C_2 \big( {\rm MS}(n) \big) 
- 2 (n+4) 
+ C_2 \big( {\rm MS}(1) \big) \big] \, |n-1,n-1,n+1\ra_{{\rm mS}}
\nonumber \\
& = \left( 3 n +7 \big) \big(n+1 \right)\, |n-1,n-1,n+1\ra_{{\rm mS}} \,.
\end{align}
In summary,
the quadratic Casimir for the MS and mS representations are
\begin{align}
C_2 \big( {\rm MS}(n+1) \big) & 
= 3 \, (n+5)(n+1) 
= 4 \, \tilde{C}_2 \big( {\rm MS}(n+1) \big)\,,
\nonumber \\
C_2 \big( {\rm mS}(n+1) \big) & = (3n+7) (n+1) 
= 4 \, \tilde{C}_2 \big( {\rm mS}(n+1) \big)\,,
\end{align}
so we have
\begin{align}
& \hspace{-0.55cm}
\xi_{ {\rm MS} (n+1)}
= \frac{1}{\,2\,}
\big[\,
\tilde{C}_2 \big( {\rm MS}(n+1) \big) 
- \tilde{C}_2 \big( {\rm MS}(n) \big)  
- \tilde{C}_2 (1) 
\, \big] 
= \frac{\, 3n \,}{4}\,, 
\nonumber \\
& \hspace{-0.55cm}
\xi_{ {\rm mS} (n+1)}
= \frac{1}{\,2\,}
\big[\,
\tilde{C}_2 \big( {\rm mS}(n+1) \big) 
- \tilde{C}_2 \big( {\rm MS}(n) \big) 
- \tilde{C}_2 (1) 
\, \big] 
= - \, \frac{\, n +4 \,}{4} \,.
\label{xi}
\end{align}
Now we are ready to evaluate
all the Casimirs for the MS($n$) representation.

\subsection{The third order Casimir and its explicit form}
\label{app;C3}

Using $\xi_{ {\rm MS} (n+1)}$ and $\xi_{ {\rm mS} (n+1)}$,
the cubic Casimir for the MS($n$) is
\begin{align}
\hspace{-0.5cm}
C_3 \big( {\rm MS}(n) \big) 
=
4^3 \tilde{C}_3 \big( {\rm MS}(n) \big) 
&= 4^3 \left[
 \left( \xi_{ {\rm MS}(n+1) } \right)^3
\frac{\, n + 4 \,}{ n+1 } 
+
\left( \xi_{ {\rm mS}(n+1) } \right)^3
\frac{\, 3n  \,}{\, n+1 \,} 
\right]
\nonumber \\
&= 24\, n(n+4)(n-2) 
\,.
\label{C3}
\end{align}
where we used 
$\tilde{C}_3= X_a X_b X_c \tr[X^a X^b X^c]
= X_a X_b X_c \tr[X_a X_b X_c]/4^3
= C_3/4^3$.

The contents of the cubic Casimir operator
are the following.
Apparently, there are
$R^3$, $\tau^3$, $\sigma^3$, $R^2 \tau$, $R^2\sigma$, 
$R \tau^2$, $R \sigma^2$, $R\tau \sigma$,
$\tau^2 \sigma$, $\tau \sigma^2$ terms,
but nonzero traces
come only from 
$R^3$, $\tau^3$, 
$\sigma^3$, $R^2 \tau$, $R^2\sigma$, and $R\tau \sigma$
terms.
We write
\begin{equation}
C_3 = C(R^3) + C(\tau^3) + C(\sigma^3)
+ C(R^2 \tau) + C(R^2 \sigma) 
+ C(R \tau \sigma) \,.
\end{equation}
$C(R^3)$ is given by
\begin{equation}
C(R^3) 
= \tr\left[ R_{ai} R_{bj} R_{ck} \right]
R_{ai} R_{bj} R_{ck} 
= - 4 \, \epsilon_{abc} \epsilon_{ijk} R_{ai} R_{bj} R_{ck} \,.
\end{equation}
$C(\tau^3)$ is\footnote{Note 
that the trace runs over $SU(4)$ fundamental 
representations so a factor $2$ should be multiplied to
the results of the $SU(2)$ fundamental representation.}
\begin{equation}
C(\tau^3) 
= \tr\left[ \tau_{a} \tau_{b} \tau_{c} \right]
\tau_{a} \tau_{b} \tau_{c} 
= 4\rmi \, \epsilon_{abc} \tau_{a} \tau_{b} \tau_{c} 
= - 8 \, \tau_a^2 \,,
\end{equation}
and $C(R^2 \tau)$ is
\begin{align}
\hspace{-0.8cm}
C(R^2 \tau) 
&= 
\tr\left[ R_{ai} R_{bj} \tau_{c} \right]
R_{ai} R_{bj} \tau_{c} 
+
\tr\left[ R_{ai} \tau_{c} R_{bj} \right]
R_{ai} \tau_{c} R_{bj}  
+
\tr\left[ \tau_{c} R_{ai} R_{bj} \right]
\tau_{c} R_{ai} R_{bj} 
\nonumber \\
&
= 4 \rmi \epsilon_{abc}
\big( 
R_{ai} R_{bj} \tau_c - R_{ai} \tau_c R_{bj} + \tau_c R_{ai} R_{bj}
\big)
\nonumber \\
&
= 4 \rmi \epsilon_{abc} R_{ai} [ R_{bj} , \tau_c ]
+ 
2\rmi \epsilon_{abc} \tau_c [ R_{ai}, R_{bj} ]
\nonumber \\
&
= - 8 \, \left( 2 R_{ai}^2 + 3 \tau_a^2 \right) \,.
\end{align}
We can obtain $C(\sigma^3)$ and $C(R^2 \sigma)$
terms as well.
Finally $C(R\tau\sigma)$ term is
\begin{align}
\hspace{-0.8cm}
C(R \tau \sigma) 
&= 
\tr\left[ R_{ai} \tau_{b} \sigma_{j} \right]
R_{ai} \tau_{b} \sigma_j 
+
\tr\left[ R_{ai} \sigma_{j} \tau_{b} \right]
R_{ai} \sigma_{j} \tau_{b}
+
\tr\left[ \tau_{b} R_{ai} \sigma_{j} \right]
\tau_{b} R_{ai} \sigma_j 
\nonumber \\
&~~
+
\tr\left[ \sigma_{j} R_{ai} \tau_{b} \right]
\sigma_{j} R_{ai} \tau_{b}
+
\tr\left[ \tau_{b} \sigma_{j} R_{ai} \right]
\tau_{b} \sigma_j R_{ai}
+
\tr\left[ \sigma_{j} \tau_{b} R_{ai} \right]
\sigma_{j} \tau_{b} R_{ai}
\nonumber \\
&
= 4 \, \big(  
2 R_{ai} \tau_{a} \sigma_i + 2 \tau_{a} \sigma_i R_{ai} 
+ \tau_{a} R_{ai} \sigma_i + \sigma_i R_{ai} \tau_{a} 
\big)
\nonumber \\
& 
= 24 R_{ai} \tau_a \sigma_i \,,
\end{align}
where we used $[R_{ai},\tau_a]=[R_{ai},\sigma_i]=0$
in the last line
to change the ordering.
Assembling all these terms,
we have
\begin{equation}
C_3
= - 4 \, \epsilon_{abc} \epsilon_{ijk} R_{ai} R_{bj} R_{ck}
- 32 C_2
+ 24 R_{ai} \tau_a \sigma_i \,.
\label{casimir3formula} 
\end{equation}
%

\section{Computations of various matrix elements}

\subsection{$\tau_a^2$ and $\sigma_i^2$
for the hedgehog state}
\label{some}

We compute $\tau_a^2$ and $\sigma_i^2$ for the hedgehog
wavefunction.
For the third component of the isospin,
we have
\begin{equation}
\la H|  \tau_3^2 | H \ra
= \frac{1}{\, 2^{\Nc} \,}
\sum_{n= 0}^{\Nc}  {}_{\Nc} C_n\, (\Nc-2n)^2
= \Nc 
\,. 
\end{equation}
For the other isospin operators, we have
\begin{align}
\la H| \left( \tau_1^2 + \tau_2^2 \right)| H \ra
& = 
\frac{1}{2} \,
\la H|  
\left\{ 
E_{ \alpha^{(1)} } - E_{ \alpha^{(2)} }\,,
E_{- \alpha^{(1)} } - E_{ -\alpha^{(2)} }
 \right\} 
| H \ra
\nonumber \\
& =
\frac{1}{2} \,
\la H|  
\left( 
E_{- \alpha^{(1)} } E_{ \alpha^{(1)} } 
+ E_{ \alpha^{(2)} } E_{ -\alpha^{(2)} }
 \right) 
| H \ra
\,, 
\end{align}
where we used the fact that
$E_{-\alpha^{(1)} } |H \ra = E_{\alpha^{(2)} } |H \ra 
= \la H | E_{-\alpha^{(2)} } = \la H | E_{\alpha^{(1)} } =0$. 
(This is because $\vec{\alpha}^{(1)} = (2,0,2)$,
$\vec{\alpha}^{(2)} = (2,0,-2)$,
and the hedgehog state has the minimum of the axial charge,
$R_{33} = -\Nc$.)
Let us further note that
\begin{align}
& \la H|  
\left( 
E_{- \alpha^{(1)} } E_{ \alpha^{(1)} } 
+ E_{ \alpha^{(2)} } E_{ -\alpha^{(2)} }
\right) 
| H \ra
\nonumber \\
=& \,
\la H|  
\left( 
[ E_{- \alpha^{(1)} } , E_{ \alpha^{(1)} } ] 
+ [ E_{ \alpha^{(2)} } , E_{ -\alpha^{(2)} } ]
\right) 
| H \ra
=
\la H|  
(\vec{\alpha}^{(2)} - \vec{\alpha}^{(1)} )_i H_i  
| H \ra
\nonumber \\
=&\,
- 4 \, \la H | R_{33} | H \ra 
= 4 \Nc \,.
\end{align}
Applying the same manipulation to the spin operators,
the hedgehog state has the following expectation values,
\begin{equation}
\la H| \,  \tau_a^{ 2}  | H \ra
= 3 \Nc 
\,, ~~~~~
\la H| \,  \sigma_i^{ 2}  | H \ra
= 3 \Nc \,.
\label{tausquare}
\end{equation}
%
\subsection{The expectation value of the 
$R\tau \sigma$ operator}
\label{Rts}

Next we compute $R_{ai} \tau_a \sigma_i$
in $C_3$ for various wavefunctions.
First we write it as
\begin{align}
R_{ai} \tau_a \sigma_i
&
= R_{33} \tau_3 \sigma_3
+ 2 \left( 
R_{+3} \tau_- \sigma_3 + R_{-3} \tau_+ \sigma_3
+ R_{3+} \tau_3 \sigma_- + R_{3-} \tau_3 \sigma_+ 
\right)
\nonumber \\
& ~~
+ 4 \left( 
R_{++} \tau_- \sigma_- + R_{-+} \tau_+ \sigma_-
+ R_{+-} \tau_- \sigma_+ + R_{--} \tau_+ \sigma_+ 
\right) \,.
\end{align}
We first consider the conventional and dichotomous wavefunctions
for the state $|p\!\up\ra$.
The isospin and spin parts can be computed readily,
and we have
\begin{equation}
\la p \! \up \!| R_{ai} \tau_a \sigma_i | p \! \up \ra
= 9 \, \la p \! \up \!| R_{33} | p \! \up \ra \,,
\end{equation}
where we used 
$ \la p\!\up\!| R_{++} | n\!\down\ra
=  \la p\!\up\!| R_{+3} | n\!\up\ra
=  \la p\!\up\!| R_{3+} | p\!\down\ra
= \la p\!\up\!| R_{33} | p\!\up\ra$,
and $\tau_- | p\! \up \ra = | n \!\up \ra$, etc.

For the hedgehog wavefunction,
it is useful to note that
$(\tau_a^{(q)} + \sigma_i^{ (q) } ) |H\ra =0$
for each $q$.
Then, for instance,
\begin{equation}
\la H| R_{33} \tau_3 \sigma_3 |H\ra
= - \Nc \la H| \tau_3 \sigma_3 |H\ra
= \Nc \la H| \tau_3^2 |H\ra = \Nc^2 \,,
\end{equation}
or another example is
\begin{equation}
\la H| R_{++} \tau_- \sigma_- | H\ra
= - \sum_q \la H| ( \tau_+^{(q)} )^2 \tau_- \sigma_- | H\ra =0\,,
\end{equation}
since $\tau_+^{(q)}$ is the fundamental representation.
Doing similar manipulations,
finally we arrive at
\begin{equation}
\la H | R_{ai} \tau_a \sigma_i | H \ra
= - \Nc^2 \,.
\end{equation}
%

\end{document}